\documentclass{article}
\usepackage{amsmath}
\usepackage{graphicx}
\usepackage[colorlinks=true,linktocpage=true,
linkcolor=blue,citecolor=blue]{hyperref}
\usepackage[a4paper]{geometry}
\begin{document}
\large
\title{\bf{One-loop QCD thermodynamics in a strong homogeneous and static 
magnetic field}}
\author{Shubhalaxmi Rath\footnote{srath.dph2015@iitr.ac.in} and Binoy Krishna 
Patra\footnote{binoyfph@iitr.ac.in}\vspace{0.1in}\\ 
Department of Physics,\\ 
Indian Institute of Technology Roorkee, Roorkee 247667, India}
\date{}
\maketitle
\begin{abstract}
We have studied how the equation of state of thermal QCD with
two light flavours is modified
in strong magnetic field by calculating the thermodynamic 
observables of hot QCD matter up to one-loop, where 
the magnetic field affects mainly the quark contribution 
and the gluonic part is largely unaffected except for the softening 
of the screening mass due to the strong magnetic field.
To  begin with the effect of magnetic field on the thermodynamics, 
we have first calculated the pressure of a thermal QCD medium
in strong magnetic field limit (SML), where the pressure at 
fixed temperature increases with the magnetic field faster than the 
increase with the temperature at constant magnetic field. This
can be envisaged from the dominant scale of thermal medium in SML, 
which is the magnetic field, like the temperature in thermal medium
in absence of strong magnetic field. Thus although the presence of strong
magnetic field makes the pressure of hot QCD medium harder
but the increase of pressure with respect to the temperature becomes 
less steeper. Corroborated to the above observations, the entropy density 
is found to decrease with the temperature in the ambience of
strong magnetic field which resonates with the fact that 
the strong magnetic field restricts the dynamics of quarks 
in two dimensions, hence the phase space gets squeezed resulting 
the reduction of number of microstates. Moreover the energy density is seen 
to decrease and the speed of sound of thermal QCD medium is increased in 
the presence of strong magnetic field.
These crucial findings in strong magnetic field could have phenomenological
implications in heavy ion collisions because the expansion dynamics of
the medium produced in noncentral ultrarelativistic heavy ion 
collisions is effectively controlled by both the energy density 
and the speed of sound.

\end{abstract}\

\newpage
\tableofcontents

\newpage
\section{Introduction}
The last three decades had witnessed hectic activities towards 
recreating the conditions similar to those existing 
shortly after the big bang, known as Quark-gluon Plasma (QGP),
in the terrestrial laboratory by carrying out collisions of ultra-relativistic 
heavy ions RHIC, BNL and LHC at  CERN, where only few events are 
truly head-on, indeed most occur 
under a finite impact parameter or centrality. As a consequence, the 
two highly charged ions impacting with a small offset may produce extremely 
large magnetic fields $ \sim m_{\pi}^2$ ($\simeq 10^{18}$ Gauss) at RHIC 
and $\sim 15 m_{\pi}^2$ at LHC \cite{Skokov:IJMP24'2009,Bzdak:PLB710'2012}.
Naive classical estimates of the lifetime of these 
magnetic fields indicate that they only exist for a 
small fraction of the lifetime of QGP \cite{Kolb:PRC67'2003, QGP}. 
However depending on the transport coefficients of the medium, the magnetic 
field may be near to its maximum strength and also be 
stationary \cite{Tuchin:PRC82'2010} - \cite{Fukushima:PRD86'2012} in its 
lifetime. Moreover the magnetic field may be assumed uniform
because even though the spatial distribution of the magnetic field is globally 
inhomogeneous but in the central region of the overlapping nuclei 
the magnetic field in the transverse plane varies very smoothly, 
which is found in the simulations of hadron-string-dynamics 
model \cite{Homogeneous} for 
${\rm{Au}}$-${\rm{Au}}$ collisions at $\sqrt{s_{NN}}=$ 200 GeV with impact 
parameter, b=10 fm. Therefore, it is worthwhile to investigate
the QCD physics in a strong homogeneous and static magnetic field, such as 
chiral magnetic effect related to the generation of electric current 
parallel to magnetic field due to difference in number 
of right and left-handed quarks \cite{KDH} - \cite{Kharzeev:PPNP75'2014}, 
axial magnetic effect due to the energy flow by 
the axial magnetic field \cite{Braguta:PRD89'2014, MAAKM}, chiral 
vortical effect due to an effective magnetic field in rotating QGP~\cite{DD, Kharzeev:PPNP88'2016}, magnetic catalysis and inverse magnetic catalysis
at finite temperature arising due to the breaking and restoration of chiral 
symmetry \cite{VVI} - \cite{AFA}, thermodynamic
properties~\cite{Andersen:JHEP1210'2012} - \cite{MVD}, 
refractive indices and decay
constant~\cite{Fayazbakhsh:PRD86'2012, Fayazbakhsh:PRD88'2013} 
of mesons in hot magnetized medium, conformal 
anomaly and production of soft photons~\cite{Basar:PRL109'2012, AJCL} at 
RHIC and LHC, dispersion relation 
in a magnetized thermal QED~\cite{Sadooghi:PRD92'2015}, synchrotron 
radiation~\cite{Tuchin:PRC87'2013}, dilepton production from 
both weakly~\cite{Tuchin:PRC87'2013} - \cite{Sadooghi:AP376'2017} and strongly \cite{Mamo:JHEP1308'2013} coupled plasma.

A variety of studies of the effects of strong magnetic fields on 
QCD thermodynamics 
have been recently resurrected by the possibility to achieve the 
magnetic fields at RHIC and LHC.
Since the magnetic field breaks the translational invariance 
in space, so the pressure becomes anisotropic 
arising due to the difference between the pressures that are 
transverse and longitudinal to the direction of background magnetic 
field, which is illustrated for an ensemble of spin one-half particles 
\cite{MVD}. Recent lattice QCD calculations~\cite{GFGSA} delve into the effects
of background magnetic fields on the equation of state (EoS) by
calculating the thermodynamic observables, namely transverse and longitudinal 
pressure, magnetization, energy density, entropy density etc. 
and inferred that the transition temperature gets reduced by the magnetic 
field \cite{Agasian:PLB663'2008} - \cite{Ayala:JPCS720'2016}.
For the hadronic matter too, the phase 
structure and the phase transitions in strong magnetic fields and 
zero quark chemical potentials have been reviewed in \cite{JWA}
through the low-energy effective theories and models, where
the thermodynamic quantities are also found to increase 
with the magnetic field~\cite{AANA}. Thus the EoS is expected to be changed
due to the magnetic field and this fascinates us to study and explore 
the modification of the EoS in presence of strong magnetic field.
In the present work, we also aim to study the thermal QCD equation of state 
perturbatively up to one-loop in a background of strong and  
homogeneous magnetic field.

For thermal medium the free energy of 
non-abelian gauge theories has been analytically 
computed up to $\mathcal{O}\left(g^4\right)$ in 
\cite{PC,PC1} ($g$, the coupling constant) and up 
to $\mathcal{O}\left(g^5\right)$ 
in \cite{CB}. The values of pressure obtained by the 
addition of successive higher order
contributions oscillate haphazardly and seem to diverge, hence
the reorganization of perturbative expansion of thermodynamic 
quantities becomes necessary. In this process, various renormalization scales 
and effective field theory methods have been emerged, such as evaluation of   
free energy by the finite temperature effective field theory methods~\cite{EA},
where the contributions coming from various 
momentum scales, {\em viz.} $T$, $gT$, and $g^2T$ are separated 
in weak-coupling regime. However, in presence of magnetic field, a thermal
medium possesses an additional scale related to the magnetic field and
depending on the strength of magnetic field compared
to the temperature of thermal medium and the quark masses, 
QCD thermodynamics has been studied in two scenarios : 
In weak field limit ($T^2 \gg |q_fB|$, $T^2 \gg m^2_f$, where $m_f$ 
and $|q_f|$ are the mass and the absolute charge of the 
quark with flavor $f$, respectively), the temperature remains the dominant 
scale of the system, so the hard thermal loop (HTL) perturbation theory 
remains the best theoretical tool in calculating the free energy 
of hot quark gluon plasma \cite{1JEM} - \cite{ANM}. 
On the other hand, in strong magnetic field limit ($|q_fB|\gg T^2$, 
$|q_fB|\gg m^2_f$), we calculate the thermodynamic observables 
by replacing the upper limit of loop momentum by the magnetic field,
which is the most prominent scale available now (precisely $\sqrt{eB}$),
like the temperature in thermal medium in absence of strong magnetic field 
in HTL perturbation theory.

In presence of magnetic field, the quark momentum $\mathbf{p}$ is 
separated into components transverse and longitudinal to the 
direction of magnetic field (say, $z$-direction),
hence the dispersion relation for quarks is modified quantum mechanically 
into
\begin{eqnarray}
E_n(p_z)=\sqrt{p_z^2+m_f^2+2n\left|q_fB\right|}
\quad,\end{eqnarray}
where $n=0$,$1$,$2$,$\cdots$ are the quantum numbers 
specifying the Landau levels. 
In strong magnetic field, the quarks are rarely excite thermally to the higher 
Landau levels, only the lowest Landau levels (LLL) $(n=0)$ are populated
($E_0=\sqrt{p_z^2+m_f^2}$). Thus the dynamics of quarks are effectively 
restricted to $(1+1)$ 
dimensions. In addition, the quark propagator is also modified in the 
magnetic field, which was first derived in
coordinate space by Schwinger~\cite{JS} using the proper-time method and 
later by Tsai~\cite{W} in the momentum space. 
With the modifications discussed above in strong magnetic
field, our aim will be to calculate the one-loop contribution to 
the thermodynamic observables of a hot strongly magnetized 
QCD matter to analyze the behaviour of QGP phase in strong magnetic field. 

Our work proceeds in the following way. In section 2, we have derived the 
effective quark propagator in a thermal QCD medium in strong magnetic field 
limit. For that purpose, we first revisit the vacuum quark propagator in the 
presence of a strong magnetic field and obtain both the quark and gluon 
propagators at finite temperature in real-time formalism (RTF), in
sections 2.1 and 2.2, respectively. This helps us to compute the one-loop 
quark self-energy in section 2.3 at finite temperature in strong magnetic 
field limit. Similarly we have derived the effective gluon propagator 
in the similar environment in section 3, where we first calculate
the one-loop gluon self-energy in terms of screening mass in section 3.1. The 
effect of magnetic field enters through the screening mass, so
we have calculated the screening mass in strong magnetic field limit
in section 3.2 for both massless and physical quark masses 
by the static limit of real part of gluon self-energy. Having thus 
obtained the effective propagators for quarks and gluons in sections
2 and 3, respectively, we have calculated the quark and gluon free energies
and then, the thermodynamic observables for a strongly magnetized 
QCD matter have been calculated in section 4. Finally we conclude in section 5.

\section{Thermalized effective quark propagator in a strongly-magnetized hot QCD medium}
The effective quark propagator in a strongly-magnetized 
hot QCD medium is obtained from the Schwinger-Dyson equation :
\begin{eqnarray}
S^{-1}(P) &=& S^{-1}_0(P)-\Sigma(P)
\quad,\label{defeffquarkprop}\end{eqnarray}
where $S_0(P)$ and $\Sigma(P)$ are the free propagator 
and quark self-energy in a strongly-magnetized hot QCD 
medium, respectively. As mentioned earlier, the strong magnetic field 
affects the quark propagator {\em via} the projection operator
and the dispersion relation, which will, in turn affect
the quark self-energy. In addition, the QCD coupling will now run with 
both the magnetic field and temperature, however, in strong magnetic 
field limit ($eB \gg T^2$), it runs exclusively with the magnetic field and is
almost independent of the temperature, because the most dominant
scale available is the magnetic field, not the temperature of medium 
anymore. For this purpose 
we closely follow the results in \cite{Ferrer:PRD91'2015},
where the coupling is split into terms dependent on the momentum parallel 
and perpendicular to the magnetic field, separately. In our case of 
magnetic field ($\mathbf{B}=B\hat{z}$), we will use the coupling which 
depends on the longitudinal component only, because
the energies of Landau levels for quarks in SML depend only on the longitudinal
component of momentum. In fact, the coupling dependent on the transverse
momentum does not depend on magnetic field at all, thus the
relevant coupling is given by~\cite{Ferrer:PRD91'2015}
\begin{equation}
\alpha_{s}^\|(eB)=\frac{g^2}{4\pi}=\frac{1}{{\alpha_s^0(\mu_0)}^{-1}+\frac{11N_c}{12\pi}
\ln\left(\frac{\Lambda_{QCD}^2+M^2_B}{\mu_0^2}\right)+\frac{1}{3\pi}\sum_f \frac{|q_f B|}{\tau}}
~,\end{equation}
where
\begin{equation}
\alpha_s^0(\mu_0) = \frac{12\pi}
{11N_c\ln\left(\frac{\mu_0^2+M^2_B}{\Lambda_V^2}\right)}
~,\end{equation}
$M_B$ is taken $\sim~1$ GeV as an infrared mass and the string tension
is taken as $\tau=0.18 ~{\rm{GeV}}^2$.

We now first revisit the vacuum quark 
propagator in a strong magnetic field and then thermalize both quark and 
gluon propagators in a hot QCD medium, which are the ingredients to
compute the quark self-energy.

\subsection{Vacuum propagators in strong magnetic field}
The magnetic field breaks the translational 
invariance of space, as a result the quark propagator becomes 
function of separate components of momentum transverse 
and longitudinal to the magnetic field direction. 
Schwinger's proper-time method \cite{JS} computes 
the quark propagator in coordinate-space as
\begin{equation}\label{S(X,Y)}
S(x,y)=\phi(x,y)\int\frac{d^4K}{(2\pi)^4}e^{-iK(x-y)}S(K)\quad,
\end{equation}
where the phase factor $\phi(x,y)$ is expressed as
\begin{equation}
\phi(x,y)=e^{i|q_f|\int^x_y A^\mu(\zeta)d\zeta_\mu}.
\end{equation}
The above phase factor is the gauge-dependent part and is 
responsible for breaking of translational invariance. 
In a single fermion propagator, 
for a symmetric gauge, i.e. $A^\mu(x)=\frac{B}{2}(0,-x_2,x_1,0)$ 
in a magnetic field directed along the $z$ axis 
($\mathbf{B}=B\hat{z}$), it is possible to gauge away the phase 
factor by an appropriate gauge transformation and one can work with the 
momentum-space representation 
of the propagator \cite{AJMMR} as an integral over the proper-time ($s$)
\begin{eqnarray}
\nonumber{S(K)} &=& i\int^{\infty}_0{ds}e^{-is{m_f}^2}\exp\left(isk_\parallel^2
-\frac{ik_\perp^2\tan(|q_fBs|)}{|q_fB|}\right) \\ && \times\left[\left(m_f+
\gamma^\parallel\cdot{k}_\parallel\right)
\left(1+\gamma^1\gamma^2\tan(|{q_fBs}|)\right)
-\gamma^\perp\cdot{k}_\perp\left(1+\tan^2(|q_fBs|)\right)\right].
\end{eqnarray}
The quantities in above equation are defined as follows
\begin{eqnarray}
&&k_\parallel\equiv(k_0,0,0,k_3),~~ k_\perp\equiv(0,k_1,k_2,0),\nonumber\\
&&\gamma^\parallel\equiv(\gamma^0,~~\gamma^3),~~ 
\gamma^\perp\equiv(\gamma^1,\gamma^2),\nonumber\\ 
&&g^{\mu\nu}=g^{\mu\nu}_\parallel+g^{\mu\nu}_\perp, \nonumber\\ 
&&g^{\mu\nu}_\parallel={\rm{diag}}(1,0,0,-1),~~
g^{\mu\nu}_\perp={\rm{diag}}(0,-1,-1,0), \nonumber\\
&&\gamma^\parallel\cdot{k}_\parallel=\gamma^0
k_0-\gamma^3k_3,~~ \gamma^\perp\cdot{k}_\perp=\gamma^1k_1+\gamma^2k_2,
\nonumber\\  
&&k^2_\parallel\equiv{k}^2_0-k^2_3,~~k^2_\perp\equiv{k}^2_1+k^2_2.\nonumber
\end{eqnarray}
After integration over the proper-time, $s$, $S(K)$ can be written
in discrete notation
\begin{equation}\label{S(K)}
S(K)=ie^{-\frac{k^2_\perp}{|q_fB|}}\sum^\infty_{n=0}
(-1)^n\frac{D_n(|q_fB|,K)}{k^2_\parallel-m^2_f-2|q_fB|n}
\quad,\end{equation}
where $D_n(|q_fB|,K)$ can be expressed in terms of generalized 
Laguerre polynomials labelling the Landau levels \cite{W,AKD,NK}
as
\begin{eqnarray}\label{D_n(|q_fB|,K)}
\nonumber{D_n(|q_fB|,K)} &=& \left(\gamma^\parallel\cdot{k}_\parallel
+m_f\right)\left[\left(1-i\gamma^1\gamma^2\right)L_n\left(\frac{2k^2_\perp}{|q_fB|}\right)-\left(1+i\gamma^1\gamma^2\right)L_{n-1}\left(\frac{2k^2_\perp}
{|q_fB|}\right)\right] \\ && +4\gamma^\perp\cdot{k}_\perp{L^{(1)}_{n-1}}
\left(\frac{2k^2_\perp}{|q_fB|}\right)
\quad.\end{eqnarray}
In presence of a strong magnetic field ( $k^2_\parallel,k^2_\perp\ll|q_fB|$),
the transitions to the higher Landau levels ($n\geq1$) are 
suppressed and only LLL ($n=0$) is occupied. 
Putting $n=0$ in equation ($\ref{D_n(|q_fB|,K)}$) 
yields the quark propagator in strong magnetic field in momentum-space
\begin{equation}\label{Q.P. in S.M.F.A.}
S_{LLL}(K)=ie^{-\frac{k^2_\perp}{|q_fB|}}\frac{\left(\gamma^\parallel
\cdot{k}_\parallel+m_f\right)}{k^2_\parallel-m^2_f}\left(1
-\gamma^0\gamma^3\gamma^5\right).
\end{equation}
However, the gluons remain unaffected by the presence 
of magnetic field, hence the form of the vacuum gluon propagator 
remains the same even in presence of magnetic field.

\subsection{Thermalization of propagators in strong magnetic field}
The vacuum quark and gluon propagators in presence of 
strong magnetic field discussed above get thermalized in
a thermal QCD medium using the RTF, where the medium effects are 
conceived through the distribution functions. In this formalism, 
propagators are manifestly separated into 
vacuum and thermal parts and the degrees of freedom has also
been doubled so the propagators acquire a $2\times{2}$ 
matrix structure. We also note that, in an 
equilibrium system, to evaluate the real part of 
the one-loop quark self energy, it is adequate 
to calculate the 11-components of the quark and gluon propagators.
\subsubsection{Quark propagator}
Since quarks are only populated in the lowest Landau levels in strong magnetic
field, so LLL quark propagator in equation (\ref{Q.P. in S.M.F.A.}) is used in the thermalization 
process. Denoting $S_{LLL}(K) =S_0(K)$, the vacuum quark propagator gets
matrix structure in thermal medium as
\begin{eqnarray}\label{Q.P.}
S(K)=U_F(k_0)\begin{pmatrix}S_0(K) & 0 \\ 0 & S^*_0(K)\end{pmatrix}U_F(k_0)
\quad,\end{eqnarray}
where $U_F(k_0)$ is the unitary matrix which 
brings the temperature dependence through the 
distribution function and it has the following 
form.
\begin{eqnarray}
U_F(k_0)=\begin{pmatrix}\sqrt{1-n_F(k_0)} & -\sqrt{n_F(k_0)} 
\\ \sqrt{n_F(k_0)} & \sqrt{1-n_F(k_0)}\end{pmatrix}
\quad,\end{eqnarray}
with the distribution function for quarks
\begin{eqnarray}
n_F(k_0)=\frac{1}{e^{\beta|k_0|}+1}
\quad.\end{eqnarray}
Substituting the unitary matrix in equation 
(\ref{Q.P.}) and simplifying, we get
\begin{eqnarray}
S(K)=\begin{pmatrix}n^2_2S_0(K)-n^2_1S^*_0(K) & -n_1n_2(S_0(K)+S^*_0(K)) 
\\ n_1n_2(S_0(K)+S^*_0(K)) & n^2_2S^*_0(K)-n^2_1S_0(K)\end{pmatrix}
\quad,\end{eqnarray}
where $n_1=\sqrt{n_F(k_0)}$ and $n_2=\sqrt{1-n_F(k_0)}$.

Now from the above matrix, the 11 - component 
of the quark propagator in a strongly 
magnetized thermal medium is obtained as
\begin{eqnarray}\label{11 Q.P.}
\nonumber{S_{11}}(K) &=& ie^{-\frac{k^2_\perp}{|q_fB|}}\left(\gamma^0k_0-\gamma^3k_3+m_f\right)\left(1
-\gamma^0\gamma^3\gamma^5\right) \\ && \times\left[\frac{1}
{k_{\parallel}^2-m_f^2+i\epsilon}+2\pi{i}n_F(k_0)
\delta\left(k_{\parallel}^2-m_f^2\right)\right],
\end{eqnarray}
which is found to be modified by the strong magnetic field.

\subsubsection{Gluon propagator}
For gluon, the unitary matrix needed to thermalize the 
propagator is of the form
\begin{eqnarray}
U_B(q_0)=\begin{pmatrix}\sqrt{1+n_B(q_0)} & \sqrt{n_B(q_0)} 
\\ \sqrt{n_B(q_0)} & \sqrt{1+n_B(q_0)}\end{pmatrix}
\quad,\end{eqnarray}
with the distribution function for gluons
\begin{eqnarray}
n_B(q_0)=\frac{1}{e^{\beta|q_0|}-1}
\quad.\end{eqnarray}
In matrix form, the gluon propagator is expressed as
\begin{eqnarray}
D^{\mu\nu}(Q)=U_B(q_0)\begin{pmatrix}D^{\mu\nu}_0(Q) & 0 \\ 0 & D^{*\mu\nu}_0(Q)\end{pmatrix}U_B(q_0).
\end{eqnarray}
Now proceeding like the quark case, the 11 - component of the gluon 
propagator can be read from the above matrix as 
\begin{eqnarray}\label{11 G.P.}
D^{\mu\nu}_{11}(Q)=ig^{\mu\nu}\left[\frac{1}{Q^2+i\epsilon}
-2\pi{i}n_B(q_0)\delta\left(Q^2\right)\right]
.\end{eqnarray}

\subsection{One-loop quark self-energy in strongly magnetized medium}
Using Feynman rules and 11-components of the 
quark and gluon propagators (equations (\ref{11 Q.P.}) 
and (\ref{11 G.P.})), the one-loop quark self energy (in figure 1) in presence 
of a strong magnetic field is given as
\begin{eqnarray}\Sigma(P) = -\frac{4}{3} g^2~i\int{\frac{d^4K}{(2\pi)^4}}\left[\gamma_\mu
{S_{11}(K)}\gamma^\mu{D_{11}(P-K)}\right]\quad,\end{eqnarray}
where the factor $4/3$ is associated with the fundamental 
representation of ${SU(3)}_c$ gauge group through the relation: 
$C_F=\frac{N^2_c-1}{2N_c}$ and $g$ is the running coupling constant.
\begin{figure}[h]
\includegraphics[width=4.9cm]{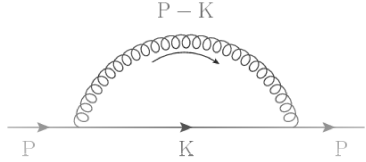}
\centering
\caption{Quark self energy}
\end{figure}

The momentum integration can be factorized into parallel and 
perpendicular components with respect to the direction of magnetic field
\begin{eqnarray}
\Sigma(P) &=& \frac{4g^2i}{3(2\pi)^4}
\int{d^2k_{\perp}}{d^2k_{\parallel}e^{-\frac{k_{\perp}^2}{\mid{q_fB}\mid}}}\left[\gamma_\mu\left(\gamma^0k_0-\gamma^3k_3
+m_f\right)\left(1-\gamma^0\gamma^3\gamma^5\right)\gamma^\mu\right] \nonumber\\
&& \times\left[\frac{1}{k_{\parallel}^2-m_f^2+i\epsilon}+2\pi{i}n_F(k_0)\delta
\left(k_{\parallel}^2-m_f^2\right)\right] \nonumber\\ && 
\times\left[\frac{1}{(P-K)^2+i\epsilon}-2\pi{i}n_B(p_0-k_0)\delta
\left((P-K)^2\right) \right]~.
\label{qsefirst}
\end{eqnarray}
In strong magnetic field limit, the external quark momentum $P$ can be 
assumed to be purely longitudinal \cite{VA}, i.e. $p_\perp=0$, so
the internal gluon momentum squared becomes 
\begin{eqnarray}(P-K)^2=\left(p_\parallel-k_\parallel\right)^2
-k_\perp^2
~.\end{eqnarray}
In LLL approximation, $k_\parallel^2$, $k_\perp^2$ are assumed to
be much smaller than $|q_fB|$, hence, the form 
factor $e^{-\frac{k_{\perp}^2}{|q_fB|}}$ can be set equal
to $1$ and the upper limits of all the momenta integrals should 
be cut off at $|q_fB|$. By evaluating the product of gamma matrices 
in (\ref{qsefirst}) as
\begin{eqnarray}\gamma_\mu\left(\gamma^0k_0-\gamma^3k_3+m_f\right)\left(1
-\gamma^0\gamma^3\gamma^5\right)\gamma^\mu=-2\left(\gamma^0k_0
-\gamma^3k_3-2m_f\right)~,
\end{eqnarray}
the quark self energy can thus be separated into the vacuum and medium 
components as
\begin{eqnarray}\label{Total sigma}
\nonumber\Sigma(p_{\parallel}) &=& \frac{-8g^2i}{3(2\pi)^4}\int{d^2k_{\perp}{d^2k_\parallel}}\left(\gamma^0k_0-\gamma^3k_3-2m_f\right) 
\\ && \nonumber\times\left[\frac{1}{k_{\parallel}^2-m_f^2+i\epsilon}+2\pi{i}n_F(k_0)\delta(k_{\parallel}^2-m_f^2)\right] 
\\ && \nonumber\times\left[\frac{1}{(p_\parallel-k_\parallel)^2
-k_\perp^2+i\epsilon}-2\pi{i}n_B(p_0-k_0)
\delta((p_\parallel-k_\parallel)^2-k_\perp^2)\right] \\ 
&\equiv& \Sigma_V(p_{\parallel})+\Sigma_n(p_{\parallel})
+\Sigma_{n^2}(p_{\parallel}),
\end{eqnarray}
where $\Sigma_V(p_{\parallel})$ represents the 
quark self-energy in vacuum, $\Sigma_n(p_{\parallel})$ 
is the quark self energy in a medium containing 
both quark and gluon distribution functions 
and $\Sigma_{n^2}(p_{\parallel})$ represents the quark self 
energy in a medium containing product of quark 
and gluon distribution functions.
\subsubsection{Vacuum part}
The vacuum contribution to the one-loop quark self energy 
is given by
\begin{eqnarray}
\nonumber\Sigma_V(p_{\parallel})=\frac{-8g^2i}{3(2\pi)^4}
\int{d^2k_{\perp}{d^2k_\parallel}}\left(\gamma^0k_0-\gamma^3k_3
-2m_f\right)\\\times\left[\frac{1}{k_{\parallel}^2-m_f^2
+i\epsilon}\right]\left[\frac{1}{(p_\parallel-k_\parallel)^2
-k_\perp^2+i\epsilon}\right]
.\end{eqnarray}
Separating the (real) principal value and imaginary part by the 
identity:
\begin{equation}\label{identity}
\frac{1}{x\pm{y}\pm{i\epsilon}}=\rm{P}\left(\frac{1}
{x\pm{y}}\right)\mp{i\pi{\delta(x\pm{y})}}
~,\end{equation}
we obtain the real part of vacuum contribution, $\Sigma_V(p_{\parallel})$ 
with the following form
\begin{eqnarray}\label{$Sigma_v$}
\Sigma_V(p_{\parallel})=\frac{-8g^2i}{3(2\pi)^4}
\int{d^2k_{\perp}d^2k_{\parallel}}\left(\gamma^\parallel
\cdot{k}_\parallel-2m_f\right)\left[\frac{1}{k_{\parallel}^2
-m_f^2}\right]\left[\frac{1}{(p_\parallel-k_\parallel)^2-k_\perp^2}\right].
\end{eqnarray}
Using the Feynman parametrization and Wick rotation, 
the parallel momentum ($k_\parallel$) integration
can be recast into the form
\begin{eqnarray}\label{K.P. integration}
I &=& i\int^1_0~dz~d^2k^\prime~\frac{z\gamma^\parallel\cdot{p_\parallel}-2m_f}{\left[{k^\prime}^2-z(1-z)p^2_\parallel+zk^2_\perp+(1-z)m^2_f\right]^2}  \nonumber\\
&=& -\frac{i\pi}{|q_fB|}\int^1_0dz\left(z\gamma^\parallel
\cdot{p}_\parallel-2m_f\right)+i\pi\int^1_0dz\frac{z\gamma^\parallel
\cdot{p}_\parallel-2m_f}{-z(1-z)p^2_\parallel+zk^2_\perp+(1-z)m^2_f} 
\nonumber\\ 
&\equiv& I_1+I_2
\quad,\end{eqnarray}
where $I_1$ is solved as
\begin{eqnarray}
\nonumber{I_1} &=& -\frac{i\pi}{|q_fB|}\int^1_0dz\left(z\gamma^\parallel
\cdot{p}_\parallel-2m_f\right) \\ 
&=& -\frac{i\pi}{2|q_fB|}\left(\gamma^\parallel\cdot{p}_\parallel-4m_f\right).
\end{eqnarray}
For solving $I_2$, we expand it in a Taylor series around the mass-shell 
condition : $\gamma^\parallel\cdot{p}_\parallel=m_f$ in a strong magnetic
field,
\begin{eqnarray}
I_2=A+B\left(\gamma^\parallel\cdot{p}_\parallel-m_f\right)
+C\left(\gamma^\parallel\cdot{p}_\parallel-m_f\right)^2+\cdots
\end{eqnarray}
Dropping the higher-order terms, the integral, $I_2$ is given by
\begin{eqnarray}\label{I_2}
I_2=A+B\left(\gamma^\parallel\cdot{p}_\parallel-m_f\right),
\end{eqnarray}
where $A$ and $B$ are given by the following expressions
\begin{eqnarray}
A &=& \left.I_2\right\vert_{\gamma^\parallel
\cdot{p}_\parallel=m_f} \nonumber\\ 
&=& i\pi{m_f}\int^1_0dz\frac{z-2}{m^2_f(1-z)^2+zk^2_\perp} \\
B &=& \left.\frac{\partial{I_2}}{\partial\left(\gamma^\parallel 
\cdot{p}_\parallel\right)}\right
\vert_{\gamma^\parallel\cdot{p}_\parallel=m_f} \nonumber\\
&=& {i\pi}\int^1_0dz\frac{z}{(1-z)^2m^2_f+zk^2_\perp}+i\pi\int^1_0dz
\frac{2m^2_fz(1-z)(z-2)} {\left((1-z)^2 m^2_f+zk^2_\perp\right)^2}
~,\end{eqnarray}
respectively. At least, for light flavours we may drop terms proportional 
to $m^2_f$ and higher orders in the solutions of above integrals to get 
$A$ and $B$ as
\begin{eqnarray}\label{A}
A &=&\frac{i\pi}{2m_f}\ln\left(\frac{k^2_\perp}{m^2_f}\right), \\
B&=&-\frac{i\pi}{2m^2_f}\ln\left(\frac{k^2_\perp}{m^2_f}\right).
\end{eqnarray}
Substituting the values for A and B in equation (\ref{I_2}), it yields
\begin{eqnarray}
\nonumber{I_2} &=& \frac{i\pi}{2m_f}\ln\left(\frac{k^2_\perp}{m^2_f}\right)-i\pi\left(\gamma^\parallel\cdot{p}_\parallel-m_f\right)
\frac{1}{2m^2_f}\ln\left(\frac{k^2_\perp}{m^2_f}\right) \\ &=& \frac{i\pi}{m_f}\ln\left(\frac{k^2_\perp}{m^2_f}\right)-\left(\gamma^\parallel\cdot{p}_\parallel\right)\frac{i\pi}{2m^2_f}\ln\left(\frac{k^2_\perp}{m^2_f}\right)
~.\end{eqnarray}
Now the integral involving the $k_\parallel$ 
integration (\ref{K.P. integration}) yields
\begin{eqnarray}
I &=& -\frac{i\pi}{2|q_fB|}\left(\gamma^\parallel\cdot{p}_\parallel
-4m_f\right)+\frac{i\pi}{m_f}\ln\left(\frac{k^2_\perp}{m^2_f}\right)-\left(\gamma^\parallel\cdot{p}_\parallel\right)\frac{i\pi}{2m^2_f}\ln\left(\frac{k^2_\perp}{m^2_f}\right)
~.\end{eqnarray}
Finally inserting the integral, $I$ in $\left(\ref{$Sigma_v$}\right)$ 
and then performing the remaining $k_\perp$ integration, we get the real
part of the vacuum contribution of one-loop quark self-energy,
\begin{eqnarray}\label{QSE in vacuum}
\Sigma_V(p_{\parallel}) 
&=& \nonumber\frac{\left(\gamma^\parallel
\cdot{p}_\parallel\right)g^2}{6\pi^2}\left[-\frac{1}{2}
-\frac{|q_fB|}{2m^2_f}\left\lbrace\ln\left(\frac{|q_fB|}{m^2_f}\right)
-1\right\rbrace\right] \\ && +\frac{g^2}{6\pi^2}\left[2m_f
+\frac{|q_fB|}{m_f}\left\lbrace\ln\left(\frac{|q_fB|}{m^2_f}\right)-1\right\rbrace\right]
.\end{eqnarray}
\subsubsection{Medium part}
In a medium, both $\Sigma_{n}(p_{\parallel})$ and $\Sigma_{n^2}(p_{\parallel})$,
which contain single quark and gluon distribution and product of quark
and gluon distribution functions, contribute to the one-loop quark 
self-energy. Now, using the identity (\ref{identity}), we obtain 
the real-part of one-loop quark self energy due to single distribution 
function, 
\begin{eqnarray}\label{S.D.F.P}
\nonumber\Sigma_n(p_{\parallel}) &=& \frac{8g^2}{3(2\pi)^3}\int{d^2k_{\perp}dk_3dk_0}\left(\gamma^0k_0
-\gamma^3k_3-2m_f\right) \\ && \nonumber\times\left[\frac{\delta
\left(k_{\parallel}^2-m_f^2\right)n_F(k_0)}{(p_\parallel-k_\parallel)^2-k_\perp^2}+\frac{\delta\left(\left(p_\parallel-k_\parallel\right)^2
-k_\perp^2\right)\left[-n_B(p_0-k_0)\right]}{k_\parallel^2-m^2_f}\right] 
\\ &\equiv& \Sigma_{n_F}(p_{\parallel})+\Sigma_{n_B}(p_{\parallel}),
\end{eqnarray}
with the quark and gluon contributions are now separable and the quark part 
is
\begin{eqnarray}
\Sigma_{n_F}(p_{\parallel}) &=& \frac{8g^2}{3(2\pi)^3}
\int{d^2k_{\perp}dk_3dk_0}\left(\gamma^0k_0-\gamma^3k_3
-2m_f\right)\frac{\delta\left(k_0^2-\omega^2_k\right)n_F(k_0)}
{(p_\parallel-k_\parallel)^2-k_\perp^2}
~,\end{eqnarray}
with $\omega^2_k=k^2_3+m^2_f$. 
After performing the $k_0$ integration using the property of 
Dirac delta function, we find
\begin{eqnarray}\label{Sigma pp.}
\Sigma_{n_F}(p_{\parallel}) &=& \frac{-8g^2}{3(2\pi)^3}
\int{dk_3}\frac{n_F\left(\omega_k\right)}{\omega_k}
\left(\gamma^3k_3+2m_f\right)\int{d^2k_\perp}\frac{1}
{\left(p_\parallel-m_f\right)^2-k^2_\perp}
~,\end{eqnarray}
which involves two independent integrations over
$k_3$ and $k_\perp$ momenta. The integral involving
$k_3$ integration is solved into~\cite{LR}
\begin{eqnarray}\label{$k_3$ integration}
I_{k_3} &=& \int^{+\infty}_{-\infty}{dk_3}
\frac{n_F\left(\omega_k\right)}{\omega_k}\left(\gamma^3k_3+2m_f\right) 
\nonumber\\ 
&=&4m_f\left[-\frac{1}{2}\ln\left(\frac{m_f}{\pi{T}}\right)-\frac{1}{2}\gamma_E+\mathcal{O}\left(\frac{m^2_f}{T^2}\right)\right]
,\end{eqnarray}
where $\gamma_E$ is the Euler-Mascheroni constant. For a thermal medium
considered here, $m^2_f$ for light flavours is much less than $T^2$, so
the term, 
$\mathcal{O}\left({m^2_f}/{T^2}\right)$ can be dropped and $I_{k_3}$ turns 
out to be
\begin{eqnarray}
I_{k_3}=-2m_f\left[\ln\left(\frac{m_f}{\pi{T}}\right)+\gamma_E\right].
\end{eqnarray}
Now the $k_\perp$ integration in equation $\left(\ref{Sigma pp.}\right)$ 
is performed after taking the upper limit of the integration by $|q_fB|$
compatible to LLL approximation
\begin{eqnarray}
I_{k_\perp} &=& -\pi\left[i\pi+\ln\left(\frac{|q_fB|}
{\left(p_\parallel-m_f\right)^2}\right)\right].
\end{eqnarray}
Substituting the values of $I_{k_3}$ and $I_{k_\perp}$ 
integrations in equation (\ref{Sigma pp.}) and 
keeping the real-part only, we obtain
\begin{eqnarray}\label{QSE in medium}
\Sigma_{n_F}(p_{\parallel}) &=& -\frac{2g^2m_f}{3\pi^2}
\ln\left(\frac{|q_fB|}{\left(p_\parallel-m_f\right)^2}\right)
\left[\ln\left(\frac{m_f}{\pi{T}}\right)+\gamma_E\right].
\end{eqnarray}
Similarly the part involving gluon distribution in equation 
$\left(\ref{S.D.F.P}\right)$ is
\begin{eqnarray}
\nonumber\Sigma_{n_B}(p_{\parallel}) &=& -\frac{8g^2}{3(2\pi)^3}
\int{d^2k_{\perp}dk_3dk_0}\left(\gamma^0k_0-\gamma^3k_3-2m_f\right) 
\\ && \times\frac{\delta\left(\left(p_\parallel-k_\parallel\right)^2
-k_\perp^2\right)}{k^2_0-\omega^2_k} n_B(p_0-k_0).
\end{eqnarray}
Simplifying the argument of Dirac delta function in the above integration 
for small $k_\perp$, 
the $k_0$ and $k_\perp$ integrations have been facilitated to yield the
contribution of gluon distribution as
\begin{eqnarray}
\nonumber\Sigma_{n_B}(p_{\parallel}) &=& -\frac{4\pi|q_fB|g^2}{3(2\pi)^3}
\int^{+\infty}_{-\infty}{dk_3}~\frac{n_B(p_3-k_3)}{\left(p_3-k_3\right)}
\left[\frac{\gamma^0p_0-2m_f}{\left(p_0+p_3-k_3\right)^2-\omega^2_k}\right. \\ && \left.\nonumber+\frac{\gamma^0\left(p_3-k_3\right)}{\left(p_0+p_3-k_3\right)^2
-\omega^2_k}-\frac{\gamma^3k_3}{\left(p_0+p_3-k_3\right)^2-\omega^2_k}+\frac{\gamma^0p_0-2m_f}{\left(p_0-p_3+k_3\right)^2
-\omega^2_k}\right. \\ && \left.-\frac{\gamma^0\left(p_3-k_3\right)}{\left(p_0-p_3+k_3\right)^2-\omega^2_k}-\frac{\gamma^3k_3}{\left(p_0-p_3+k_3\right)^2
-\omega^2_k}\right].
\end{eqnarray}
Finally the above $k_3$ integration can be integrated out to give
\begin{eqnarray}
\Sigma_{n_B}(p_{\parallel})=-\frac{4\pi|q_fB|g^2}{3(2\pi)^3}\left(I^1+I^2+I^3+I^4+I^5+I^6\right)
,\end{eqnarray}
where $I^1$, $I^2$, $I^3$, $I^4$, $I^5$ and $I^6$ are
found as follows
\begin{eqnarray}
&&I^1=\frac{i\pi\left(\gamma^0{p_0}-2m_f\right)}{2\left(p_0+p_3\right)}
\left[\frac{\beta(a-p_3)-2}{2\beta(p_3-a)^2}+\frac{n_B(a-p_3)}{a-p_3}\right]~, \\
&&I^2=\frac{-i\pi\gamma^0}{2\left(p_0+p_3\right)}
\left[\frac{1}{\beta(p_3-a)}+n_B(a-p_3)\right]~, \\
&&I^3=\frac{-i\pi\gamma^3}{2\left(p_0+p_3\right)}
\left[\frac{-2a-\beta{p_3}(p_3-a)}{2\beta(p_3-a)^2}+\frac{an_B(a-p_3)}{a-p_3}\right]~, \\
&&I^4=\frac{-i\pi\left(\gamma^0{p_0}-2m_f\right)}{2\left(p_0-p_3\right)}
\left[\frac{\beta(b-p_3)-2}{2\beta(p_3-b)^2}+\frac{n_B(b-p_3)}{b-p_3}\right]~, \\
&&I^5=\frac{-i\pi\gamma^0}{2\left(p_0-p_3\right)}
\left[\frac{1}{\beta(p_3-b)}+n_B(b-p_3)\right]~, \\
&&I^6=\frac{i\pi\gamma^3}{2\left(p_0-p_3\right)}
\left[\frac{-2b-\beta{p_3}(p_3-b)}{2\beta(p_3-b)^2}+\frac{bn_B(b-p_3)}{b-p_3}
\right]~,\end{eqnarray}
where $a$ and $b$ are given by
\begin{eqnarray}
&&a=\frac{\left(p_0+p_3\right)^2-m^2_f}{2\left(p_0+p_3\right)}~, \\
&&b=\frac{\left(p_0-p_3\right)^2-m^2_f}{2\left(p_3-p_0\right)}
~.\end{eqnarray}
Therefore, $\Sigma_{n_B}(p_{\parallel})$ cannot contribute to the 
real part of one-loop quark self-energy.

Finally, the medium contribution to the quark self 
energy involving product of quark and gluon distribution functions 
(from equation $(\ref{Total sigma})$) is
\begin{eqnarray}
\nonumber\Sigma_{n^2}(p_{\parallel}) &=& \frac{-8g^2i}{3(2\pi)^4}
\int{d^2k_{\perp}{d^2k_\parallel}}4{\pi^2}n_F(k_0)n_B(p_0-k_0)\left(\gamma^0k_0-\gamma^3k_3-2m_f\right) 
\\ && \times\left[\delta(k_{\parallel}^2-m_f^2)
\delta((p_\parallel-k_\parallel)^2-k_\perp^2)\right],
\end{eqnarray}
which however does not contribute to the real part of 
quark self energy. 

Thus the vacuum (\ref{QSE in vacuum}) and the medium contributions (\ref{QSE 
in medium}) are added together to give the real part of one-loop quark 
self energy of a thermal QCD medium in strong magnetic field
\begin{eqnarray}\label{Q.S.E.}
\nonumber\Sigma(p_{\parallel}) &=& \frac{\left(\gamma^\parallel
\cdot{p}_\parallel\right)g^2}{6\pi^2}\left[-\frac{1}{2}
-\frac{|q_fB|}{2m^2_f}\left\lbrace\ln\left(\frac{|q_fB|}{m^2_f}\right)
-1\right\rbrace\right] \\ && \nonumber+\frac{g^2}{6\pi^2}
\left[2m_f+\frac{|q_fB|}{m_f}\left\lbrace\ln\left(\frac{|q_fB|}
{m^2_f}\right)-1\right\rbrace\right] \\ && -\frac{2g^2m_f}{3\pi^2}
\ln\left(\frac{|q_fB|}{\left(p_\parallel-m_f\right)^2}\right)
\left[\ln\left(\frac{m_f}{\pi{T}}\right)+\gamma_E\right],
\end{eqnarray}
which enables us to compute the effective quark propagator from Dyson-Schwinger equation 
(\ref{defeffquarkprop}).

\section{Thermalized effective gluon propagator in a strongly-magnetized hot QCD medium}
This section is attributed to the evaluation 
of effective gluon propagator in a thermal 
medium in presence of a strong magnetic field. 
In general, the effective gluon propagator can 
be obtained from the Schwinger-Dyson equation
\begin{eqnarray}
D^{-1}_{\mu\nu}(P)=D^{-1}_{0\mu\nu}(P)+\Pi_{\mu\nu}(P)
~.\end{eqnarray}
At finite temperature, the effective gluon propagator gets 
decomposed into longitudinal and transverse 
components in thermal medium. Although gluons are
not affected directly by the presence of magnetic 
field but the dependence of magnetic field enters directly through the 
Debye mass and indirectly through the running strong coupling. To evaluate the components 
of gluon propagator, first we revisit how to decompose the gluon self-energy 
in a thermal medium in the coming subsection.

\subsection{One-loop gluon self-energy in a hot QCD medium}
In vacuum, the gluon self-energy tensor is the linear
combination of available four-momentum of  the particle
($P_\mu$) and the metric tensor $(g_{\mu \nu})$. Being a
Lorentz invariant quantity, self-energy depends on $P^2$ and further
restriction by the Ward identity : $P^\mu\Pi_{\mu\nu}(P)=0$ imposes
the structure of the tensor as
\begin{eqnarray}
\Pi_{\mu\nu}(P)&=&\left(g_{\mu \nu} - \frac{P_\mu{P_\nu}}{P^2}\right)
\Pi(P^2)\nonumber\\
&\equiv &P_{\mu \nu} \Pi(P^2),
\end{eqnarray}
where $P_{\mu\nu}$ is the transverse projection operator. However, at finite 
temperature, the Lorentz invariance is broken 
due to the direction of heat bath, which is
introduced in terms  of a four-velocity, $u_\mu$ in the rest frame of 
heat bath. Now with the available four vectors, $P_\mu$, $u_\mu$ and 
the tensor, $g_{\mu\nu}$, two orthogonal tensors, which are most adopted
to the physical degrees of freedom and project
on the subspace transverse and parallel to the three momentum, $\mathbf{p}$, 
respectively, are constructed
\begin{eqnarray}
P^T_{\mu\nu}&=&g_{\mu \nu}-\frac{P_\mu P_\nu}{P^2} -\frac{P^L_{\mu \nu}}
{-N^2}~, \\
P^L_{\mu\nu}&=& -N_\mu N_\nu, {\rm{with}}~~ N_\mu=
\frac{P_\mu(P.u)-u_\mu P^2}{{(P.u)}^2-P^2}
~,\end{eqnarray}
as the tensorial basis to decompose the gluon self-energy tensor in 
thermal medium
\begin{eqnarray}\label{G.S.E.}
\Pi_{\mu\nu}(p_0,\mathbf{p})=P^T_{\mu\nu}\Pi_T(p_0,\mathbf{p})+P^L_{\mu\nu} 
\Pi_L(p_0,\mathbf{p})\quad.\end{eqnarray}
The above functions $\Pi_T$ and $\Pi_L$ are known as transverse and 
longitudinal self-energies, respectively, which 
depends on both energy and three momentum in the rest frame of
medium due to lack of Lorentz invariance as
\begin{eqnarray}
p_0&=&u^\mu\cdot{P_\mu}~, \\
|\mathbf{p}|&=&\sqrt{\left(u^\mu\cdot{P_\mu}\right)^2-P^2}
\quad.\end{eqnarray}
By using the properties of the above projection operators, $P^L_{\mu \nu}$, 
and $P^T_{\mu \nu}$, the 
transverse and longitudinal self energies can be obtained as
\begin{eqnarray}\label{S.T.G.S.E.}
\Pi_L(P)&=&-\Pi_{00}(P)~, \\
\Pi_T(P)&=&\frac{1}{2}\left(\Pi_\mu^\mu(P) - \frac{P^2}{p^2} \Pi_L(P)\right)
~,\end{eqnarray}
which are obtained in HTL perturbation theory~\cite{JEEM}. The HTL gluon 
self-energy tensor in a thermal medium determined
by the angular average over the spatial directions of light-like vectors is
\begin{eqnarray}
\nonumber\Pi_{\mu\nu}(P) &=& m^2_D\left[\int\frac{d\Omega}{4\pi}
K_\mu{K_\nu}\frac{P_\mu\cdot{u^\mu}}{P_\mu\cdot{K^\mu}}-u_\mu{u_\nu}\right]
\\ &=& m^2_D\left[\int\frac{d\Omega}{4\pi}
K_\mu{K_\nu}\frac{p_0}{p_0+\mathbf{p}\cdot{\mathbf{\hat{k}}}}-u_\mu{u_\nu}\right]
,\end{eqnarray}
where $K_\mu=(1,\mathbf{\hat{k}})$ is a light like
four-vector and $m_D$ is the Debye mass. Therefore, the transverse and 
longitudinal components, 
$\Pi_T(P)$ and $\Pi_L(P)$ become
\begin{eqnarray}\label{T.G.S.E.}
\Pi_T(P) &=& \frac{m^2_D}{2}\frac{p^2_0}{p^2}
+\frac{m^2_D}{4}\frac{p_0}{p}\left(1-\frac{p^2_0}{p^2}\right)
\ln\left(\frac{p_0+p}{p_0-p}\right)\\
\Pi_L(P) &=& m^2_D-\frac{m^2_D}{2}\frac{p_0}{p}
\ln\left(\frac{p_0+p}{p_0-p}\right),
\label{L.G.S.E.}
\end{eqnarray}
respectively. Thus the Schwinger-Dyson equation gives the dressed thermal 
gluon propagator as
\begin{eqnarray}\label{G.P.}
D_{\mu\nu}=P^T_{\mu\nu}~\Delta_T+P^L_{\mu \nu} \frac{P^2}{p^2} \Delta_L
~,\end{eqnarray}
where the transverse and longitudinal components of gluon propagator are given 
by
\begin{eqnarray}
\Delta_T&=&\frac{-1}{P^2+\Pi_T(P)}\label{T.G.P.}~, \\
\Delta_L&=&\frac{1}{p^2+\Pi_L(P)}\label{L.G.P.}
~.\end{eqnarray}
Physically, $\Delta_T$ describes the propagation of two transverse vacuum modes
in thermal medium whereas $\Delta_L$ does not exist in the vacuum and thus 
represents collective modes of medium. 

Now when the medium becomes strongly magnetized, the 
dependence of strong magnetic field in $\Pi_{T}(P)$ 
and $\Pi_L(P)$ originates from the magnetic field
dependence of the Debye mass. Therefore we are going to derive
the Debye mass in strong magnetic field by the static limit of
the longitudinal component of gluon self energy in the next subsection. 

\subsection{Screening mass in strong magnetic field}
The Debye screening manifests in the collective oscillation of
the medium via the dispersion relation and is obtained by the
static limit of the longitudinal part (``00'' component)
of gluon self-energy, i.e.
\begin{eqnarray}\label{Definition}
\Pi_L (p_0=0,\mathbf{p}\rightarrow 0)=m^2_D
~.\end{eqnarray}
Out of four contributing diagrams (tadpole, gluon loop, ghost loop and quark 
loop) of the gluon self energy, only quark-loop (in figure 2) gets influenced 
by the magnetic field.
\begin{figure}[h]
\includegraphics[width=4.9cm]{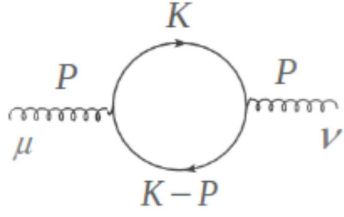}
\centering
\caption{Gluon self energy}
\end{figure}

In real-time formalism, using Schwinger's proper-time
propagator in strong magnetic field (\ref{11 Q.P.}) for internal
quark line, the 11-component of the gluon self-energy for the quark-loop 
is written as
\begin{eqnarray}
\nonumber\Pi^{\mu\nu}(P) &=& -\frac{ig^2}{2}
\int\frac{d^4K}{(2\pi)^4}tr\left[\gamma^\mu{S_{11}(K)}
\gamma^\nu{S_{11}(K-P)}\right] \\ &=& \nonumber\frac{ig^2}
{2(2\pi)^4}\sum_f\int{d^2k_\perp}{d^2k_\parallel}{tr}\left[\gamma^\mu
\left(\gamma^0k_0-\gamma^3k_3+m_f\right)\gamma^\nu
\left(\gamma^0q_0-\gamma^3q_3+m_f\right)\right] 
\\ && \nonumber\times\left[\frac{1}{k^2_\parallel-m^2_f+i\epsilon}+2\pi{i}n_F\left(k_0\right)\delta\left(k^2_\parallel-m^2_f\right)\right]e^{-\frac{k^2_\perp}{|q_fB|}} 
\\ && \times\left[\frac{1}{q^2_\parallel-m^2_f+i\epsilon}+2\pi{i}n_F\left(q_0\right)\delta\left(q^2_\parallel-m^2_f\right)\right]e^{-\frac{q^2_\perp}{|q_fB|}}
\quad,\end{eqnarray}
where the factor $1/2$ enters due to the trace 
over colour indices and we use $Q=\left(q_0,\mathbf{q}\right)$ 
in place of $K-P=\left(k_0-p_0,\mathbf{k}-\mathbf{p}\right)$. The 
momentum integration is factorized into parallel 
and perpendicular components with respect to the direction of magnetic 
field, where the $k_\perp$ integration, $\Pi_{k_\perp}(p_\perp)$ is
separated out to give
\begin{eqnarray}\label{P.C.G.S.E.}
\nonumber{\Pi_{k_\perp}(p_\perp)} &=& \int{dk_1}{dk_2}e^{-\frac{k^2_\perp}
{|q_fB|}}e^{-\frac{q^2_\perp}{|q_fB|}} \\ &=& \frac{\pi|q_fB|}{2}
e^{-\frac{p^2_\perp}{2|q_fB|}}.
\end{eqnarray}
Thus the gluon self energy becomes completely separable into the 
components of momentum which are parallel and perpendicular to the magnetic 
field as 
\begin{eqnarray}
\Pi^{\mu\nu}(P) &=& \frac{\pi|q_fB|}{2}e^{-\frac{p^2_\perp}
{2|q_fB|}}\Pi^{\mu\nu}(p_\parallel).
\end{eqnarray}
Denoting the trace over gamma matrices by $L^{\mu \nu}$ as
\begin{eqnarray}
L^{\mu\nu}=8\left[k^\mu_\parallel\cdot{q^\nu_\parallel}
+k^\nu_\parallel\cdot{q^\mu_\parallel}-g^{\mu\nu}_\parallel\left(k^\mu_\parallel\cdot{q}_{\parallel\mu}
-m^2_f\right)\right]
~,\end{eqnarray}
the self energy depending only on the parallel component of momentum is 
decomposed into vacuum and thermal contributions as
\begin{eqnarray}\label{G.S.E.S.M.F.A.}
\nonumber\Pi^{\mu\nu}(p_\parallel) &=& \frac{ig^2}
{2(2\pi)^4}\sum_f\int{dk_0dk_3}L^{\mu\nu}\left[\frac{1}
{k^2_\parallel-m^2_f+i\epsilon}+2\pi{i}n_F\left(k_0\right)
\delta\left(k^2_\parallel-m^2_f\right)\right] 
\\ && \nonumber\times\left[\frac{1}{q^2_\parallel-m^2_f
+i\epsilon}+2\pi{i}n_F\left(q_0\right)\delta
\left(q^2_\parallel-m^2_f\right)\right] 
\\ &\equiv & \Pi^{\mu\nu}_V(p_\parallel)+\Pi^{\mu\nu}_n(p_\parallel)
+\Pi^{\mu\nu}_{n^2}(p_\parallel)
\quad,\end{eqnarray}
where $\Pi^{\mu \nu}_V(p_{\parallel})$, $\Pi^{\mu \nu}_n(p_{\parallel})$ 
and $\Pi^{\mu \nu}_{n^2} (p_{\parallel})$ are the vacuum and medium 
contributions to the gluon self energy containing 
single and double distribution functions, respectively, and are given 
by 
\begin{eqnarray}
\Pi^{\mu \nu}_V(p_{\parallel}) &=& \frac{ig^2}{2(2\pi)^4}
\int dk_0 dk_3 L^{\mu\nu}\left[\frac{1}{(k_{\parallel}^2
-m_f^2+i\epsilon)}\frac{1}
{(q_{\parallel}^2-m_f^2+i\epsilon)}\right],
\label{G.S.E.V.} \\
\Pi^{\mu \nu}_n(p_{\parallel}) &=& -\frac{g^2}
{2(2\pi)^3}\int dk_0 dk_3 L^{\mu\nu}\left[\frac{n_{F}(k_0)
\delta(k_{\parallel}^2-m_f^2)}{(q_{\parallel}^2
-m_f^2+i\epsilon)}
+\frac{n_{F}(q_0)\delta(q_{\parallel}^2-m_f^2)}
{(k_{\parallel}^2-m_f^2+i\epsilon)}\right],
\label{G.S.E.S.D.} \\
\Pi^{\mu \nu}_{n^2} (p_{\parallel}) &=& -\frac{ig^2}{2(2\pi)^2}
\int dk_0 dk_3 L^{\mu\nu}\left[n_{F}(k_0)n_{F}(q_0)
\delta(k_{\parallel}^2-m_f^2)
\delta(q_{\parallel}^2-m_f^2)\right]~.
\label{G.S.E.D.D.}
\end{eqnarray}

We will now first evaluate the vacuum part (\ref{G.S.E.V.}) for which the real
part (using the identity (\ref{identity})) is given by
\begin{equation}
\Pi^{\mu\nu}_V(p_\parallel)=\left(g_{\parallel}^{\mu\nu}
-\frac{p_{\parallel}^{\mu}p_{\parallel}^{\nu}}{p_{\parallel}^2}
\right)\Pi(p_\parallel^2),
\end{equation}
where $\Pi (p_\parallel^2)$ is given by
\begin{eqnarray}
\Pi (p_\parallel^2)=\frac{g^2}{2\pi^3}\sum_{f} \left[\frac{2m_{f}^2}
{p_{\parallel}^2}
\left(1-\frac{4m_{f}^2}{p_{\parallel}^2}\right)^{-1/2}
\ln \left\lbrace \frac{{\Big(1-\frac{4m_{f}^2}
{p_{\parallel}^2}\Big)}^{1/2}+1}
{{\Big(1-\frac{4m_{f}^2}{p_{\parallel}^2}
\Big)}^{1/2}-1} \right\rbrace +1\right].
\end{eqnarray}
The real part of 00 - component of vacuum contribution to
the one-loop gluon self-energy tensor becomes
\begin{equation}
\Pi^{00}_V (p_0, p_3)=-\frac{p_{3}^2}{p_{\parallel}^2}
~\Pi (p_\parallel^2)
~.\end{equation}

Thus multiplying the transverse component (\ref{P.C.G.S.E.}) of gluon self-energy,
the vacuum part of one-loop gluon self energy for massless flavours in 
the static limit ($p_0=0$, $p_1, p_2, p_3 \rightarrow 0$) is simplified into
\begin{equation}
\Pi^{00}_V =\frac{g^2}{4\pi^2}\sum_{f}|q_{f}B| ,
\label{Massless case}
\end{equation}
whereas for physical quark masses, it vanishes in the static limit, 
\begin{equation}
\Pi^{00}_V = 0
~.\label{Massive case}
\end{equation}

Similarly the real part of $00$ - component of the thermal contribution 
to the gluon 
self-energy, $\Pi^{\mu\nu}_n(p_\parallel)$ with a single distribution 
function for $p_0=0$ is given by
\begin{eqnarray}\label{D.M.S.(p_3)}
\nonumber\Pi^{00}_n(p_0=0,p_3) &=& -\frac{g^2}
{2(2\pi)^3}\sum_f\int{dk_3}\left[\frac{L^{00}(k_0=\omega_k)n_F\left(k_0=\omega_k\right)}{2\omega_k\{\omega_k^2-\omega^2_q\}}\right. \\ && \left.\nonumber+\frac{L^{00}(k_0=-\omega_k)n_F\left(k_0=-\omega_k\right)}{2\omega_k\{-\omega_k^2-\omega^2_q\}}
+\frac{L^{00}(k_0=\omega_q)n_F\left(k_0=\omega_q\right)}{2\omega_k\{\omega_q^2-\omega^2_k\}}
\right. \\ && \left.+\frac{L^{00}(k_0=-\omega_q)n_F\left(k_0=-\omega_q\right)}{2\omega_k\{(-\omega_q)^2-\omega^2_k\}}\right]
,\end{eqnarray}
where the different factors in above equation are given by
\begin{eqnarray}
&&L^{00}=8\left(k^0q^0+k^3q^3+m^2_f\right), \nonumber\\
&&\omega^2_k=k^2_3+m^2_f,~\omega^2_q=q^2_3+m^2_f, \nonumber\\
&&L^{00}(k_0=\omega_k)=8\left(2\omega^2_k-k_3p_3\right), \nonumber\\
&&L^{00}(k_0=-\omega_k)=8\left(2\omega^2_k-k_3p_3\right), \nonumber\\
&&L^{00}(k_0=\omega_q)=8\left(\omega^2_k+\omega^2_q-k_3p_3\right), \nonumber\\
&&L^{00}(k_0=-\omega_q)=8\left(\omega^2_k+\omega^2_q-k_3p_3\right), \nonumber\\
&&n_F\left(k_0=\omega_k\right)=n_F\left(k_0=-\omega_k\right)=\frac{1}{e^{\beta|\omega_k|}+1}, \nonumber\\
&&\nonumber{n_F\left(k_0=\omega_q\right)}=n_F\left(k_0=-\omega_q\right)=\frac{1}{e^{\beta|\omega_q|}+1}
.\end{eqnarray} 
For massless quarks, the medium contribution, $\Pi^{00}_n(p_0=0,p_3)$ reduces 
to 
\begin{equation}\label{Massless}
\Pi^{00}_n(p_0=0,p_3)=\frac{8g^2}{2(2\pi)^3} 
\left[-1-\frac{T}{p_3}\ln(2)
+\frac{T}{p_3}\ln\left(1+e^{\frac{p_3}{T}}\right)\right]
,\end{equation}
whereas for the physical quark masses, it becomes
\begin{eqnarray}\label{Massive}
\Pi^{00}_n(p_0=0,p_3) &=& -\frac{g^2}{2(2\pi)^3}
\int  dk_3 \left[ \frac{8k_{3}n_F(\omega_k)}{\omega_{k}p_{3}}\right. \nonumber \\
&&  \left. -\frac{8(k_{3}-p_{3})n_F(\omega_q)}{\omega_{q}p_{3}}+\frac{16m_{f}^2n_F(\omega_k)}
{\omega_{k}p_{3}(2k_{3}-p_{3})}-\frac{16m_{f}^2n_F(\omega_q)}
{\omega_{q}p_{3}(2k_{3}-p_{3})} \right]
.\end{eqnarray}

The medium contribution to the self-energy with the square of quark 
distribution function (\ref{G.S.E.D.D.}) is purely imaginary so it 
does not have the real-part, i.e.
\begin{eqnarray}
\Pi^{\mu \nu}_{n^2}(p_{\parallel})=0
~.\end{eqnarray}

Again multiplying the transverse component (\ref{P.C.G.S.E.}) to 
parallel component (\ref{Massless}) gives the real part of 
$00$ - component of medium contribution to one-loop 
gluon self-energy tensor for massless quarks, which, in the static 
limit ($p_0=0$, $p_3 \rightarrow 0$), is simplified 
into 
\begin{equation}\label{Massless (M.C)}
\Pi^{00}_n=-\frac{g^2}{4\pi^2}\sum_{f} 
|q_{f}B|+\frac{g^2}{8\pi^2}\sum_{f}|q_{f}B|
~,\end{equation}
whereas the real part of $00$ - component 
of medium contribution to one-loop gluon self-energy tensor for physical 
quark masses in the static limit is reduced to
\begin{equation}\label{Massive (M.C.)}
\Pi^{00}_n=\frac{g^2}{4\pi^{2}T}\sum_{f}|q_{f}B|\int_{0}^{\infty}dk_{3}\frac{e^{\beta\omega_{k}}}{(1+e^{\beta\omega_{k}})^2}
~.\end{equation}

Finally the vacuum (\ref{Massless case}) and medium contributions 
(\ref{Massless (M.C)}) are 
added together to give the real part of the $00$ - component of one-loop 
gluon self-energy in static limit for massless quarks
\begin{equation}
\Pi^{00}= \frac{g^2}{8\pi^2}\sum_{f} 
|q_{f}B|
~,\label{a}
\end{equation}
whereas the vacuum (\ref{Massive case}) and medium contributions 
(\ref{Massive (M.C.)}) yield the real part of the $00$ - component of one-loop 
gluon self-energy tensor in static limit for physical quark masses
 \begin{equation}
\Pi^{00}=\frac{g^2}{4\pi^{2}T}
\sum_{f}|q_{f}B|\int_{0}^{\infty}
dk_{3}\frac{e^{\beta\omega_{k}}}{(1+e^{\beta\omega_{k}})^2}
~.\label{b}
\end{equation}
Therefore the definition (\ref{Definition}) gives the Debye mass 
for the massless quarks
\begin{equation}\label{$m_D^2$}
m_{D}^2=\frac{g^2}{8\pi^2}\sum_{f}|q_{f}B|
~,\end{equation}
which was recently obtained by one of us~\cite{MBB} and by others by different
approach~\cite{Fukushima:PRD93'2016,Bandyopadhyay:PRD94'2016}. It is found 
that the Debye mass 
of thermal QCD medium in the presence of strong magnetic field
depends solely on the magnetic field 
and is independent of temperature, therefore the collective behaviour of 
the medium gets strongly affected by the presence of 
strong magnetic field. However, for physical quark masses, the Debye mass is 
obtained as
\begin{eqnarray}\label{D.M.S.}
m^2_D=\frac{g^2}{4\pi^2T}\sum_f|q_fB|\int^\infty_0dk_3
\frac{e^{\beta\sqrt{k^2_3+m^2_f}}}{\left(1+e^{\beta\sqrt{k^2_3
+m^2_f}}\right)^2}
~,\end{eqnarray}
which depends now on both magnetic field and temperature. However 
it becomes independent of temperature beyond 
a certain temperature for a particular strong magnetic field~\cite{MBB}.

\section{Thermodynamic observables of QCD matter in strong magnetic field}

\subsection{Free energy and pressure}
The one loop free energy for $N_f$ quarks with $N_c$ 
colours in hot QCD medium in a static and homogeneous 
strong magnetic field is given 
by the sum of free energies due to quarks ($\mathcal{F}_q$) and gluons
($\mathcal{F}_g$), which are obtained 
by the functional determinant of effective
quark and gluon propagators, respectively. Finally 
the pressure for the quark matter is obtained by the 
negative of free energy in the thermodynamic limit. We are now in 
a position to calculate the free energies and hence the pressures due to
quarks and gluons using their respective one-loop propagators.

\subsubsection{Quark contribution}
The free energy due to $N_f$ quarks with $N_c$ colours is obtained
by the effective quark propagator, $S(P)$ from equation (\ref{defeffquarkprop})
\begin{eqnarray}\label{Free energy due to quarks}
\mathcal{F}_q &=& N_cN_f\int\frac{d^4P}{(2\pi)^4}\ln\left[\det\left(S(P)
\right)\right] \nonumber\\
&=&-N_cN_f\int\frac{d^4P}{(2\pi)^4}\ln\left[\det\left(\gamma^\parallel
\cdot{p}_\parallel-m_f-\Sigma(p_\parallel)\right)\right]
.\end{eqnarray}
Due to the external magnetic field in $z$ direction, the momentum integration
in the quark free energy is also factorized into the momentum parallel
and perpendicular to the magnetic field, which is facilitated by the 
finding that the integrand (i.e. the effective propagator)  
depends only on the longitudinal momentum component. For the sake of 
simplicity, we 
first express the quark self-energy into dependent and independent 
terms on (parallel) momentum from equation (\ref{Q.S.E.})
\begin{eqnarray}
\Sigma(p_\parallel)&=&\left(\gamma^\parallel
\cdot{p}_\parallel\right)C+D+E, ~{\rm{with}} \\
C&=&\frac{g^2}{6\pi^2}\left[-\frac{1}{2}
-\frac{|q_fB|}{2m^2_f}\left\lbrace\ln\left(\frac{|q_fB|}
{m^2_f}\right)-1\right\rbrace\right]\label{C}, \\
D&=&\frac{g^2}{6\pi^2}\left[2m_f+\frac{|q_fB|}{m_f}
\left\lbrace\ln\left(\frac{|q_fB|}{m^2_f}\right)-1\right\rbrace\right]\label{D}, \\
E&=&-\frac{2g^2m_f}{3\pi^2}\ln\left(\frac{|q_fB|}{\left(p_\parallel-m_f
\right)^2}\right)\left[\ln\left(\frac{m_f}{\pi{T}}\right)+\gamma_E\right]\label{E},
\end{eqnarray}
and then evaluate the determinant,
\begin{eqnarray}
\det\left[\gamma^\parallel\cdot{p}_\parallel-m_f-\Sigma(p_\parallel)\right]=\left[p^2_\parallel\left(1
-C\right)^2-\left(m_f+D+E\right)^2\right]^2.
\end{eqnarray}
Thus after plugging the determinant into the 
integration (\ref{Free energy due to quarks}), 
the thermodynamic free energy of QCD matter due to quark contribution
is expressed as
\begin{eqnarray}\label{F.E.Q.C.}
\nonumber\mathcal{F}_q &=& -2N_cN_f
\int\frac{d^2p_\perp}{(2\pi)^2}\int\frac{d^2p_\parallel}
{(2\pi)^2}\ln\left[p^2_\parallel\left(1-C\right)^2
-\left(m_f+D+E\right)^2\right] \\ &=& \nonumber-2N_cN_f
\int\frac{d^2p_\perp}{(2\pi)^2}\left[\int\frac{d^2p_\parallel}{(2\pi)^2}
\ln\left(p^2_\parallel\right)+\int\frac{d^2p_\parallel}{(2\pi)^2}\ln\left[\left(1
-C\right)^2-\frac{1}{p^2_\parallel}\left(m_f+D+E\right)^2\right]\right] 
\\ &\equiv& -\frac{N_cN_f|q_fB|}{2\pi}\left(I_{1p_\parallel}+I_{2p_\parallel}
\right)
.\end{eqnarray}
At finite temperature, the integrals of type 
$I_{1p_\parallel}$ have been frequently solved 
in $4$-dimension employing Matsubara 
frequency sum method, where the continuous energy integrals 
are replaced by discrete frequency sums. Thus we solve this analytically as
\begin{eqnarray}
\nonumber{I_{1p_\parallel}} &=& \int\frac{dp_0}{2\pi}\int\frac{dp_3}
{2\pi}\ln\left(p^2_0-p^2_3\right) \\ &=& \frac{\pi{T^2}}{6}
~.\end{eqnarray}
The integral, $I_{2p_\parallel}$ cannot 
be evaluated analytically, so we compute it numerically 
\begin{eqnarray}
\nonumber{I_{2p_\parallel}} &=& \int\frac{d^2p_\parallel}
{(2\pi)^2}\ln\left[\left(1-C\right)^2-\frac{1}{p^2_\parallel}
\left(m_f+D+E\right)^2\right] \\ 
&=& \frac{1}{4\pi}\int^{|q_fB|}_0{dp^2_\parallel}\ln\left[\left(1-C\right)^2-\frac{1}{p^2_\parallel}\left\lbrace m_f+D+W\ln\left(\frac{|q_fB|}{\left(p_\parallel-m_f\right)^2}\right)\right\rbrace^2\right]
,\end{eqnarray}
by reexpressing the (parallel) momentum-dependent term, $E$ as
\begin{eqnarray}
E &=& W\ln\left(\frac{|q_fB|}{\left(p_\parallel-m_f\right)^2}\right)~,~
{\rm{with}}\\
W &=& -\frac{2g^2m_f}{3\pi^2}\left[\ln\left(\frac{m_f}{\pi{T}}\right)
+\gamma_E\right]\label{W}
.\end{eqnarray}
Now, substituting the integrals $I_{1p_\parallel}$ and 
$I_{2p_\parallel}$ in equation (\ref{F.E.Q.C.}), the 
free energy due to quark contribution of hot QCD matter 
in strong magnetic field is obtained as
\begin{eqnarray}
\nonumber\mathcal{F}_q &=& -\frac{N_cN_f|q_fB|}{4}
\left[\frac{T^2}{3}+\frac{1}{2\pi^2}\int^{|q_fB|}_0{dp^2_\parallel}
\ln\left[\left(1-C\right)^2\right.\right. \\ 
&& \left.\left.-\frac{1}{p^2_\parallel}\left\lbrace m_f+D+W\ln\left(\frac{|q_fB|}{\left(p_\parallel-m_f\right)^2}\right)\right\rbrace^2\right]\right].
\end{eqnarray}
Hence the negative of the above free energy in the thermodynamic limit
gives the quark contribution to the thermodynamic pressure
\begin{eqnarray}
\nonumber{P_q} &=& \frac{N_cN_f|q_fB|}{4}\left[\frac{T^2}{3}
+\frac{1}{2\pi^2}\int^{|q_fB|}_0{dp^2_\parallel}\ln\left[\left(1
-C\right)^2\right.\right. \\ && \left.\left.-\frac{1}{p^2_\parallel}
\left\lbrace m_f+D+W\ln\left(\frac{|q_fB|}{\left(p_\parallel
-m_f\right)^2}\right)\right\rbrace^2\right]\right].
\end{eqnarray}
After substituting the values of $C$, $D$ and $W$ 
(equations (\ref{C}), (\ref{D}) and (\ref{W})) 
in above equation, we get the pressure due to quark contribution
\begin{eqnarray}
\nonumber{P_q} &=& \frac{N_cN_f|q_fB|}{4}\left[\frac{T^2}{3}
+\int^{|q_fB|}_0\frac{dp^2_\parallel}{2\pi^2}\ln\left[\left(1
-\frac{g^2}{6\pi^2}\left\lbrace-\frac{1}{2}
-\frac{|q_fB|}{2m^2_f}\left(\ln(\frac{|q_fB|}
{m^2_f})-1\right)\right\rbrace\right)^2\right.\right. 
\\ &-& \left.\left.\nonumber\frac{1}{p^2_\parallel}
\left(m_f+\frac{g^2}{6\pi^2}\left\lbrace2m_f+\frac{|q_fB|}{m_f}
\left(\ln(\frac{|q_fB|}{m^2_f})-1\right)\right\rbrace\right.\right.\right. 
\\ &-& \left.\left.\left.\frac{2g^2m_f}{3\pi^2}
\left\lbrace\ln\left(\frac{m_f}{\pi{T}}\right)
+\gamma_E\right\rbrace\ln\left(\frac{|q_fB|}{\left(p_\parallel
-m_f\right)^2}\right)\right)^2\right]\right].
\end{eqnarray}

\subsubsection{Gluon contribution}
The free energy due to gluons in adjoint representation of 
$SU(N_c)$ gauge theory is given by both transverse and longitudinal modes
\begin{eqnarray}\label{F.E.G.}
\nonumber\mathcal{F}_g &=& (N_c^2-1)\left[2\mathcal{F}^T_g
+\mathcal{F}^L_g\right] \\ &=& \nonumber(N_c^2-1)\left[
\int\frac{d^4P}{(2\pi)^4}\ln\left[-\Delta_T(P)\right]^{-1}
+\frac{1}{2}\int\frac{d^4P}{(2\pi)^4}\ln\left[\Delta_L(P)\right]^{-1}\right] 
\\ &=& -\left(N_c^2-1\right)\left[\int\frac{d^4P}{(2\pi)^4}\ln\left[-\Delta_T(P)\right]+\frac{1}{2}\int\frac{d^4P}{(2\pi)^4}\ln\left[\Delta_L(P)\right]\right],
\end{eqnarray}
where $\Delta_T(P)$ and $\Delta_L(P)$ are the transverse 
and longitudinal parts of the hard thermal loop gluon propagator,
respectively, obtained earlier in equations (\ref{T.G.P.},\ref{L.G.P.}). 

Then substituting the values of $\Delta_T(P)$ and $\Delta_L(P)$ 
in equation (\ref{F.E.G.}), we obtain the gluon 
contribution to the thermodynamic free energy of QCD matter 
in strong magnetic field
\begin{eqnarray}
\mathcal{F}_g &=& -\left(N_c^2-1\right)\left[
\int\frac{d^4P}{(2\pi)^4}\ln\left(\frac{1}{P^2+\Pi_T(P)}\right)\right.
\nonumber\\ &+& \left.\frac{1}{2}\int\frac{d^4P}{(2\pi)^4}\ln\left(\frac{1}{p^2
+\Pi_L(P)}\right)\right],
\end{eqnarray}
where $\Pi_L$ and $\Pi_T$ depend on the magnetic field through the
screening (Debye) mass (\ref{D.M.S.}). 

For gluon, the loop momenta may be hard (i.e. order of $T$) 
or soft (i.e. order of $gT$). In imaginary-time, the gluon 
energy $p_0$ is an integer multiple of $2\pi{T}$, so the soft region requires 
$p_0=0$. But for quark, the energy, $p_0=(2n+1)\pi{T}$ can never 
be zero even for $n=0$, so, the quark loop is always hard. 
Since the gluons are not affected by the presence of magnetic 
field, the highest scale for them in a medium is 
still the temperature. For the hard loop 
momenta, the self energy components act like 
perturbative corrections and thereby making the 
effective propagators to expand around them. In 
case of soft loop momenta, in the zero frequency (static)
mode, $\Pi_T$ becomes zero and only $\Pi_L$ exists. 
Since this is as large as the inverse of free propagator, 
the effective propagator can not be expanded. The 
total free energy due to gluon contribution is therefore 
obtained by adding the free energies determined 
from both hard and soft scales. Up to $\mathcal{O}\left(g^4\right)$ in
strong running coupling, the gluon free energy is 
calculated in~\cite{JEEM}
\begin{eqnarray}
\nonumber\mathcal{F}_g &=& -\left(N_c^2-1\right)\left[\frac{\pi^2T^4}{45}
-\frac{T^2m^2_D}{24}+\frac{Tm^3_D}{12\pi}\right. \\ && \left.+\left\lbrace\frac{1}{\epsilon}
+2\ln\left(\frac{\Lambda}{4\pi{T}}\right)-7+2\gamma_E+\frac{2\pi^2}{3}\right\rbrace
\frac{m^4_D}{128\pi^2}\right]
,\end{eqnarray}
where $\Lambda$ is the renormalization scale. The divergent term
($1/\epsilon$) is isolated by the dimensional regularization,
which is taken care of by the vacuum counter term as
\begin{eqnarray}
\Delta\mathcal{E}_0=\frac{N^2_c-1}{128\pi^2\epsilon}m^4_D
\quad.\end{eqnarray}
After adding the counter term, the renormalized gluon free energy 
for a strongly magnetized thermal QCD matter takes the following form
\begin{eqnarray}\label{$F_g$}
\nonumber\mathcal{F}_g &=& -\left(N_c^2-1\right)\left[\frac{\pi^2T^4}{45}
-\frac{T^2m^2_D}{24}+\frac{Tm^3_D}{12\pi}\right. \\ && \left.+\left\lbrace2\ln\left(\frac{\Lambda}{4\pi{T}}\right)-7+2\gamma_E+\frac{2\pi^2}{3}\right\rbrace
\frac{m^4_D}{128\pi^2}\right].
\end{eqnarray}
Hence the pressure is obtained from 
the negative of the above free energy as
\begin{eqnarray}\label{$P_g$}
\nonumber{P_g} &=& \left(N_c^2-1\right)\left[\frac{\pi^2T^4}{45}
-\frac{T^2m^2_D}{24}+\frac{Tm^3_D}{12\pi}\right. \\ && \left.
+\left\lbrace2\ln\left(\frac{\Lambda}{4\pi{T}}\right)-7
+2\gamma_E+\frac{2\pi^2}{3}\right\rbrace\frac{m^4_D}{128\pi^2}\right].
\end{eqnarray}
After plugging the Debye mass in strong magnetic field limit (\ref{$m_D^2$}) for
light flavours, the magnetic field dependence in the gluon 
contribution will be explicitly seen as  
\begin{eqnarray}\label{$P_g$}
\nonumber{P_g} &=& \left(N_c^2-1\right)\left[\frac{\pi^2T^4}{45}
-g^2\frac{T^2{eB}}{192\pi^2}+g^3\frac{T(eB)^{\frac{3}{2}}}{192\sqrt{2}\pi^4}\right. \\ && \left.
+g^4\frac{(eB)^2}{8192\pi^6}\left\lbrace2\ln\left(\frac{\Lambda}{4\pi{T}}\right)-7
+2\gamma_E+\frac{2\pi^2}{3}\right\rbrace\right].
\end{eqnarray}

\subsubsection{Total pressure}
The total one-loop pressure of hot QCD matter 
in a strong magnetic field is obtained 
by adding both quark and gluonic contributions 
and has the following expression.
\begin{eqnarray}\label{Total pressure}
\nonumber{P(T,eB)} &=& \frac{N_cN_f|q_fB|}{4}\left[\frac{T^2}{3}
+\int^{|q_fB|}_0\frac{dp^2_\parallel}{2\pi^2}\ln\left[\left(1
-\frac{g^2}{6\pi^2}\left\lbrace-\frac{1}{2}
\right.\right.\right.\right. \\ && \left.\left.\left.\left.\nonumber-\frac{|q_fB|}{2m^2_f}\left(\ln(\frac{|q_fB|}
{m^2_f})-1\right)\right\rbrace\right)^2\right.\right. 
\\ && \left.\left.\nonumber-\frac{1}{p^2_\parallel}
\left(m_f+\frac{g^2}{6\pi^2}\left\lbrace2m_f+\frac{|q_fB|}{m_f}
\left(\ln(\frac{|q_fB|}{m^2_f})-1\right)\right\rbrace\right.\right.\right. 
\\ && \left.\left.\left.-\frac{2g^2m_f}{3\pi^2}
\left\lbrace\ln\left(\frac{m_f}{\pi{T}}\right)
+\gamma_E\right\rbrace\ln\left(\frac{|q_fB|}{\left(p_\parallel
-m_f\right)^2}\right)\right)^2\right]\right] \nonumber\\
&& \nonumber+\left(N_c^2-1\right)\left[\frac{\pi^2T^4}{45} 
-\frac{T^2m^2_D}{24}+\frac{Tm^3_D}{12\pi}
\right. \\ && \left.+\left\lbrace2\ln\left(\frac{\Lambda}{4\pi{T}}\right)-7
+2\gamma_E+\frac{2\pi^2}{3}\right\rbrace\frac{m^4_D}{128\pi^2}\right]
,\end{eqnarray}
where the renormalization scale $\Lambda$ is set at $2\pi{T}$. The ideal 
component can thus be read as 
\begin{eqnarray}\label{I.P.S.M.F.A.}
P_{ideal}(T,eB)=N_cN_f\frac{|q_fB|T^2}{12}
+\left(N_c^2-1\right)\frac{\pi^2T^4}{45}
~.\end{eqnarray}

\begin{figure}[t]
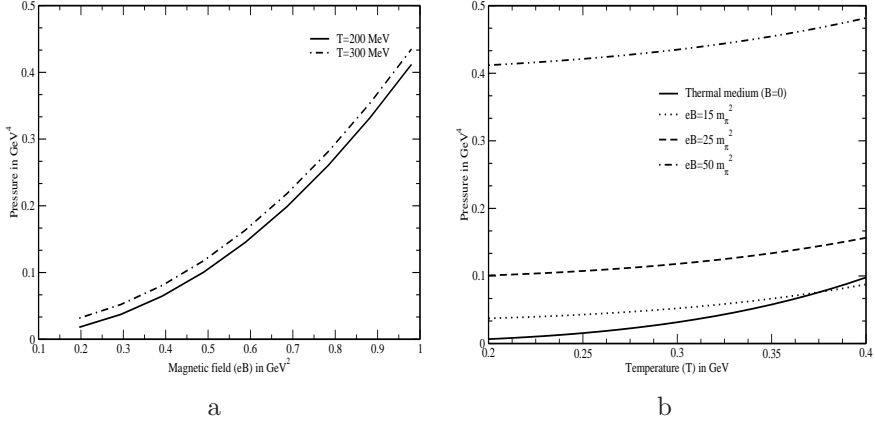

\begin{center}
\begin{tabular}{c c}
\includegraphics[width=5.5cm,height=5cm]{p3.eps}&
\includegraphics[width=5.5cm,height=5cm]{p4.eps}\\
a & b
\end{tabular}
\caption{The variation of the pressure of a hot QCD medium for
two light flavours with the strong magnetic fields at fixed temperatures
(a) and as a function of temperature at different strong magnetic fields (b).
}
\end{center}
\label{fig1}
\end{figure}
Before calculating the thermodynamic observables in the strong magnetic 
field limit, we should be careful in choosing the range of temperatures 
and magnetic fields compatible with the limit. {\em For example}, to observe 
the variation of pressure with magnetic field at a temperature $T=300$ MeV, 
the starting value of magnetic field has to be greater than 
$\sim 4.60~m_\pi^2$, however, we have taken the starting 
magnetic field, $eB=10~m_\pi^2$, which is almost twice the above marginal 
value. Similarly for calculating the 
pressure as a function of temperature up to $T=400$ MeV, 
we have fixed the magnetic fields at $eB=15~m_\pi^2$, $25~m_\pi^2$, 
and $50~m_\pi^2$.

To see how the pressure of hot QCD matter with two light flavours is affected 
by the ambient strong magnetic fields quantitatively, we have computed 
the pressure as a function of magnetic field at fixed temperatures, 
$T=200$ MeV and $300$ MeV of the medium in figure 3a, whereas the figure 3b 
denotes the 
variation of pressure with the temperature at different strong magnetic fields,
$ eB= 15$ $m^2_\pi$, $25$ $m^2_\pi$, and $50$ $m^2_\pi$.
It is found that the pressure increases rapidly with the magnetic field
(figure 3a) compared to its much slower variation with the temperature 
(figure 3b). The above contrast in the behaviour of pressure with 
magnetic field compared to the temperature reflects the 
fact that the dominant scale of thermal medium in strong magnetic field 
limit is the magnetic field, not the temperature anymore happened to be 
in thermal medium in absence of magnetic field. To be more precise, 
although the variation of pressure with temperature is not steeper
in the presence of strong magnetic field but the 
pressure for thermal QCD in strong magnetic field is 
larger than the pressure of thermal medium in absence of strong
magnetic field (denoted by solid line in figure 3b), hence the strong magnetic 
field makes the equation of state for a thermal medium harder. These
observations will facilitate in understanding the effects of strong magnetic 
field on the entropy density (figure 5). Recently, the 
thermodynamic pressure and other observables arising due to the presence 
of magnetic fields have also been studied 
extensively in lattice QCD with $2+1$ flavour \cite{GFGSA}, where they have 
noticed the similar increasing trend of the longitudinal pressure 
with the increase of magnetic field strengths.
\begin{figure}[t]
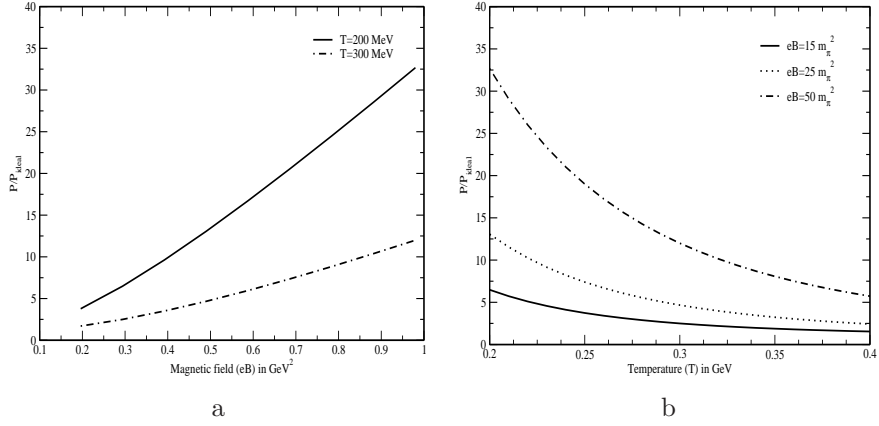

\begin{center}
\begin{tabular}{c c}
\includegraphics[width=5.5cm,height=5cm]{pr3.eps}&
\includegraphics[width=5.5cm,height=5cm]{pr4.eps}\\
a & b
\end{tabular}
\caption{The variation of QCD pressure normalized 
by its ideal value as a function of magnetic field 
for different temperatures (a) and as a 
function of temperature for different strong magnetic 
fields (b).}
\end{center}
\label{fig1}
\end{figure}
To see how the pressure of interacting quarks and gluons in a hot QCD medium
in presence of strong magnetic field approaches 
to the noninteracting (ideal) limit asymptotically both with the magnetic 
field and the temperature, we have computed the pressure in units of ideal
pressure (\ref{I.P.S.M.F.A.}) as function of magnetic fields (in figure 4a)
and temperatures as well (in figure 4b). From the figure 4a, it is 
interesting to know that 
the deviation of pressure from its ideal value increases with the
strong magnetic field, i.e. the thermal QCD medium never achieve its ideal limit
asymptotically in the presence of strong magnetic field. On
the other hand, it is found from figure 4b that, the pressure of the thermal medium 
at a fixed magnetic field approaches its ideal limit as expected, 
but in strong magnetic field limit one cannot arbitrarily increase 
the temperature due to its constraint ($eB \gg T^2$).

\subsection{Entropy density}
To see how the available microstates to a given macrostate of a thermal 
QCD medium are 
affected due to the presence of strong magnetic field, we calculate
the entropy density of hot QCD matter in a strong magnetic field 
by partially differentiating the pressure (\ref{Total pressure}) 
with respect to the temperature
\begin{eqnarray}
S&=&\frac{\partial{P}}{\partial{T}} \nonumber\\
& \equiv & S_q + S_g
~,\end{eqnarray}
where the entropy density due to quark contribution is calculated as
\begin{eqnarray}
\nonumber{S_q} &=& \frac{N_cN_f|q_fB|}{6}
\left[T-\frac{m_fg^2}{T\pi^4}\int^{|q_fB|}_0dp^2_\parallel
~\frac{1}{p^2_\parallel}\ln\left(\frac{|q_fB|}
{\left(p_\parallel-m_f\right)^2}\right)\right. 
\\ && \left.\nonumber\times\left[\left(m_f+\frac{g^2}{6\pi^2}\left\lbrace2m_f+\frac{|q_fB|}{m_f}
\left(\ln(\frac{|q_fB|}{m^2_f})-1\right)\right\rbrace\right.\right.\right. \\ && \left.\left.\left.\nonumber-\frac{2g^2m_f}{3\pi^2}\left\lbrace\ln\left(\frac{m_f}{\pi{T}}\right)
+\gamma_E\right\rbrace\ln\left(\frac{|q_fB|}
{\left(p_\parallel-m_f\right)^2}\right)\right)\right.\right. \\ && \left.\left. \nonumber\Big{/}\left(\left(1-\frac{g^2}{6\pi^2}\left\lbrace-\frac{1}{2}
-\frac{|q_fB|}{2m^2_f}\left(\ln(\frac{|q_fB|}
{m^2_f})-1\right)\right\rbrace\right)^2\right.\right.\right. \\ && \left.\left.\left.\nonumber-\frac{1}{p^2_\parallel}\left(m_f+\frac{g^2}{6\pi^2}\left\lbrace2m_f+\frac{|q_fB|}{m_f}
\left(\ln(\frac{|q_fB|}{m^2_f})-1\right)\right\rbrace\right.\right.\right.\right. \\ && \left.\left.\left.\left.-\frac{2g^2m_f}{3\pi^2}\left\lbrace\ln\left(\frac{m_f}{\pi{T}}\right)
+\gamma_E\right\rbrace\ln\left(\frac{|q_fB|}{\left(p_\parallel-m_f\right)^2}\right)\right)^2\right)\right]\right].
\end{eqnarray}
Similarly, the entropy density due to gluonic contribution 
is calculated as 
\begin{eqnarray}
{S_g} &=& \left(N^2_c-1\right)\left[\frac{4\pi^2T^3}{45}
-\frac{Tm^2_D}{12}+\frac{m^3_D}{12\pi}\right], 
\end{eqnarray}
which will be seen to depend on magnetic field after substituting the Debye mass
for two light flavours from (\ref{$m_D^2$}) as
\begin{eqnarray}
{S_g} &=& \left(N^2_c-1\right)\left[\frac{4\pi^2T^3}{45}
-g^2\frac{TeB}{96\pi^2}+g^3\frac{(eB)^{\frac{3}{2}}}{192\sqrt{2}\pi^4}\right].
\end{eqnarray}

In calculating the above expression, we set the partial derivative of
the square of screening mass with respect to the temperature to zero, i.e. 
$\frac{\partial\left(m^2_D\right)}{\partial{T}}\simeq0$, because
the screening mass in strong magnetic field limit solely
depends on the magnetic field for massless flavours, it may however
depend on both magnetic field and temperature for realistic physical
quark masses but the temperature dependence is still negligible.
\begin{figure}[t]
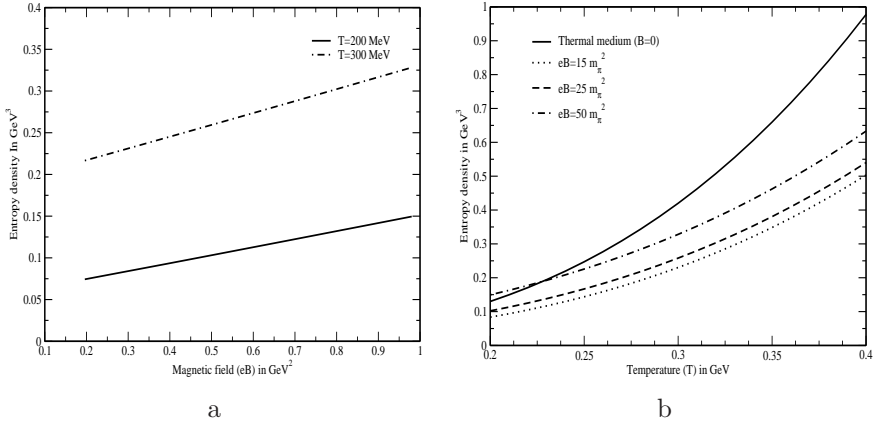

\begin{center}
\begin{tabular}{c c}
\includegraphics[width=5.5cm,height=5cm]{e1.eps}&
\includegraphics[width=5.5cm,height=5cm]{e2.eps}\\
a & b
\end{tabular}
\caption{The variation of entropy density as a function of magnetic field 
at different temperatures (a) and as a function 
of temperature at different strong  magnetic fields (b). }
\end{center}
\label{fig1}
\end{figure}

In figure 5a, we have shown how the entropy density of a hot
QCD medium at fixed temperatures, $T=200$ MeV and $300$ MeV has 
been affected by 
the stronger magnetic fields, where we found that it increases
almost linearly with the magnetic field. 
Similarly figure 5b depicts the variation of entropy density of a hot QCD 
medium with the increasing temperatures
in strong magnetic field backgrounds, $eB=15 m_\pi^2$, $25 m_\pi^2$,
and $50 m_\pi^2$. It is also seen in figure 5b that both the entropy density 
and its rate of increase with temperature in presence of strong magnetic field 
get waned compared to the medium in absence 
of strong magnetic field (denoted by the solid line, $B$=0). This
crucial observation can be understood qualitatively: As we know that 
the strong magnetic field
restricts the dynamics of quarks in momentum space from 4-dimension to 
2-dimension, hence the phase space gets shrunk to 2-dimension only.
Since the entropy is a measure of number of possible microstates in the 
phase space, so the entropy of thermal medium gets decreased in
the presence of strong magnetic field.
Thus both figures of the entropy density corroborate the observations in the 
variation of 
pressure with magnetic field and temperature (figures 3a and 3b), respectively. 

\subsection{Energy density}
We are now in a position to calculate the energy density in baryonless 
($\mu_q=0$) hot QCD medium in a strong magnetic field. This can be 
obtained from the following thermodynamic relation,
\begin{eqnarray}\label{Total energy density}
\varepsilon &=& -P+TS\nonumber\\
& \equiv& \varepsilon_q+ \varepsilon_g
~,\end{eqnarray}
where the energy density due to quark contribution is calculated as
\begin{eqnarray}
\nonumber\varepsilon_q &=& -P_q+TS_q \\ &=& \nonumber\frac{N_cN_f|q_fB|}{6}
\left[\frac{T^2}{2}-\frac{3}{4\pi^2}\int^{|q_fB|}_0{dp^2_\parallel}
~\ln\left[\left(1-\frac{g^2}{6\pi^2}\left\lbrace-\frac{1}{2}
\right.\right.\right.\right. \\ && \left.\left.\left.\left.\nonumber-\frac{|q_fB|}{2m^2_f}\left(\ln(\frac{|q_fB|}
{m^2_f})-1\right)\right\rbrace\right)^2\right.\right. \\ && \left.\left.\nonumber-\frac{1}{p^2_\parallel}\left(m_f+\frac{g^2}{6\pi^2}\left\lbrace2m_f+\frac{|q_fB|}{m_f}
\left(\ln(\frac{|q_fB|}{m^2_f})-1\right)\right\rbrace\right.\right.\right. \\ && \left.\left.\left.\nonumber-\frac{2g^2m_f}{3\pi^2}\left\lbrace\ln\left(\frac{m_f}{\pi{T}}\right)
+\gamma_E\right\rbrace\ln\left(\frac{|q_fB|}{\left(p_\parallel-m_f\right)^2}\right)\right)^2\right]\right. \\ && \left.\nonumber-\frac{m_fg^2}
{\pi^4}\int^{|q_fB|}_0dp^2_\parallel~\frac{1}{p^2_\parallel}
\ln\left(\frac{|q_fB|}{\left(p_\parallel-m_f\right)^2}\right)\right. \\ && \left.\nonumber\times\left[\left(m_f+\frac{g^2}{6\pi^2}\left\lbrace2m_f+\frac{|q_fB|}{m_f}
\left(\ln(\frac{|q_fB|}{m^2_f})-1\right)\right\rbrace\right.\right.\right. \\ && \left.\left.\left.\nonumber-\frac{2g^2m_f}{3\pi^2}\left\lbrace\ln\left(\frac{m_f}{\pi{T}}\right)
+\gamma_E\right\rbrace\ln\left(\frac{|q_fB|}
{\left(p_\parallel-m_f\right)^2}\right)\right)\right.\right. \\ && \left.\left. \nonumber\Big{/}\left(\left(1-\frac{g^2}{6\pi^2}\left\lbrace-\frac{1}{2}
-\frac{|q_fB|}{2m^2_f}\left(\ln(\frac{|q_fB|}
{m^2_f})-1\right)\right\rbrace\right)^2\right.\right.\right. \\ && \left.\left.\left.\nonumber-\frac{1}{p^2_\parallel}\left(m_f+\frac{g^2}{6\pi^2}\left\lbrace2m_f+\frac{|q_fB|}{m_f}
\left(\ln(\frac{|q_fB|}{m^2_f})-1\right)\right\rbrace\right.\right.\right.\right. \\ && \left.\left.\left.\left.-\frac{2g^2m_f}{3\pi^2}\left\lbrace\ln\left(\frac{m_f}{\pi{T}}\right)
+\gamma_E\right\rbrace\ln\left(\frac{|q_fB|}{\left(p_\parallel-m_f\right)^2}
\right)\right)^2\right)\right]\right].
\end{eqnarray}
Similarly, the energy density due to gluonic contribution has been calculated 
as 
\begin{eqnarray}
\nonumber\varepsilon_g &=& -P_g+TS_g \\ &=& \left(N^2_c-1\right)
\left[\frac{\pi^2T^4}{15}-\frac{T^2m^2_D}{24}-\left\lbrace2\ln\left(\frac{\Lambda}{4\pi{T}}\right)-7+2\gamma_E+\frac{2\pi^2}{3}\right\rbrace
\frac{m^4_D}{128\pi^2}\right].
\end{eqnarray}
The magnetic field dependence can be seen after replacing the Debye mass
for massless flavours from (\ref{$m_D^2$})
\begin{eqnarray}
\nonumber\varepsilon_g &=& \left(N^2_c-1\right)
\left[\frac{\pi^2T^4}{15}-g^2\frac{T^2{eB}}{192\pi^2}\right. \\ && \left.-g^4\frac{(eB)^2}{8192\pi^6}\left\lbrace2\ln\left(\frac{\Lambda}{4\pi{T}}\right)-7+2\gamma_E+\frac{2\pi^2}{3}\right\rbrace\right].
\end{eqnarray}

\begin{figure}[h]
\includegraphics[width=5.5cm,height=5cm]{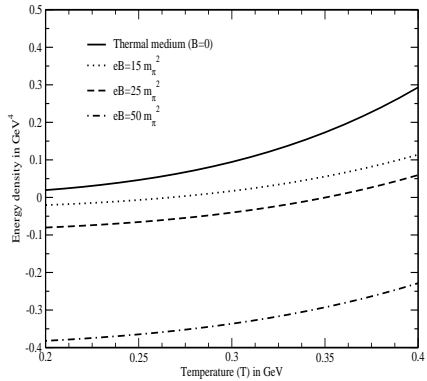}
\centering
\caption{The variation of energy density with temperature 
at different magnetic fields.}
\end{figure}

To see how the energy density of a hot QCD medium has been affected 
by the presence of external strong magnetic field, we have computed
the energy density as a function of temperature 
at different (strong) magnetic fields in figure 6, where 
the energy density increases with the temperature as expected 
but it increases with the temperature much faster for the medium in absence 
of magnetic field (B=0, denoted by solid line) which resonates with
the observation of entropy density with temperature in figure 5b. In
brief, the strong magnetic field reduces the energy density 
of thermal QCD medium.

\subsection{Speed of sound}
The speed of sound in a medium depends on the nature of equation of state, 
whether it is soft or hard and is related to the thermodynamic pressure and 
energy density through the following equation
\begin{eqnarray}
C^2_s=\frac{\partial{P}}{\partial\varepsilon}=\frac{{\partial{P}}/{\partial{T}}}
{{\partial\varepsilon}/{\partial{T}}}
~,\end{eqnarray}
where the partial derivatives of pressure and energy density with respect 
to the temperature are obtained from equations (\ref{Total pressure}) 
and (\ref{Total energy density}), respectively. Since the existence of strong 
magnetic field modifies the thermodynamic observables, 
$c^2_s$ is also expected to deviate from its value 
in the presence of strong magnetic field.
\begin{figure}[t]
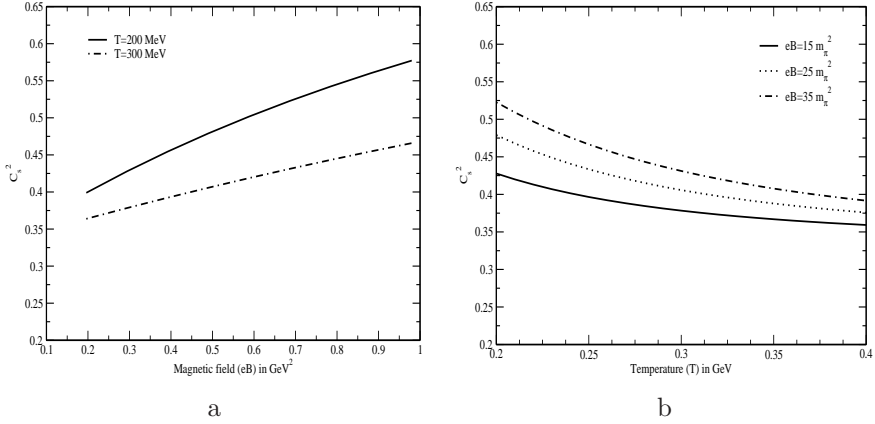

\begin{center}
\begin{tabular}{c c}
\includegraphics[width=5.5cm,height=5cm]{c1.eps}&
\includegraphics[width=5.5cm,height=5cm]{c2.eps}\\
a & b
\end{tabular}
\caption{The variation of the square of the speed of sound 
with the strong magnetic field at various temperatures (a)
and the variation with temperature in presence of strong 
magnetic fields of different strengths (b).}
\end{center}
\end{figure}
To see the effects of the magnetic fields on the equation of state, 
we have computed the speed of sound of a hot QCD medium as a function 
of external magnetic fields in figure 7a. It is
found that the speed of sound at a fixed temperature increases with the strength 
of magnetic fields, which can be understood
by the fact that, as the strength of magnetic field increases, the
energy density decreases and the pressures increases, hence the speed of 
sound gets increased. From the original perspective of how the speed of sound
of a thermal QCD is now modified in the presence of magnetic field, 
we have computed $c_s^2$ as a function of temperature at different
strengths of magnetic fields in figure 7b. We found that
$c_s^2$ decreases with the temperature as expected 
and reaches asymptotically to the ideal value 1/3 for the
case when there is no magnetic field. 

Interestingly when we plot the speed of sound in terms of ideal 
limit in strong magnetic field in figure 8, we have found that the
speed of sound shows a dip in specific temperature-magnetic field
combination. The above crucial observations in strong magnetic field 
could have phenomenological implications in heavy ion collisions, 
because the speed of sound modulates the hydrodynamic expansion of 
the medium (QGP) produced in noncentral ultrarelativistic heavy ion 
collisions.
\begin{figure}[t]
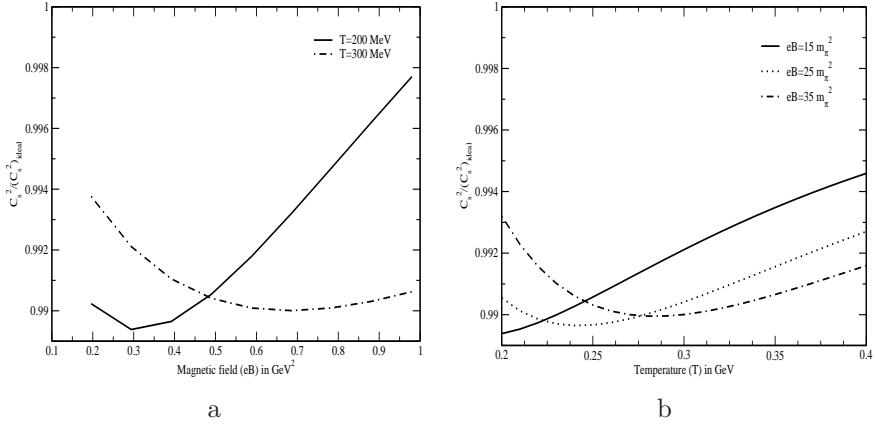

\begin{center}
\begin{tabular}{c c}
\includegraphics[width=5.5cm,height=5cm]{cr1.eps}&
\includegraphics[width=5.5cm,height=5cm]{cr2.eps}\\
a & b
\end{tabular}
\caption{The variation of $c_s^2$ normalized by its ideal value as a 
function of magnetic field at various temperatures (a) 
and as a function of temperature in presence 
of varying strong magnetic field strengths (b).}
\end{center}
\end{figure}
\section{Conclusions}
In this work, we have explored how the thermodynamic observables of
a hot QCD medium in one-loop have been affected in an ambience of
very strong magnetic field, which may be produced in noncentral events 
of ultrarelativistic
heavy ion collisions. All thermodynamic observables have been
contributed both by quarks and gluons through their respective
one-loop self energies, where the quark
contribution has been affected strongly by the strong magnetic field
whereas, the gluonic part
is largely unaffected except for the softening of the screening mass
in strong magnetic field. As a result, even the pressure for the
noninteracting quarks in thermal medium gets enhanced in strong 
magnetic field and overall an increase in total pressure of 
thermal medium is observed compared to the thermal medium
in the absence of strong magnetic field. As a consequence, 
the entropy density gets decreased due to the presence of strong
magnetic field, so the energy density too decreases with respect
to pure thermal medium. Finally we obtain the equation of
state by calculating the speed of sound of thermal QCD medium, which 
is found to increase due to the presence of strong magnetic field and shows 
a dip in a specific range of temperatures and strong magnetic fields.
The above observations could have interesting implications on the
expansion dynamics of the medium produced at RHIC and LHC in presence
of strong magnetic field.
\section{Acknowledgements}
We convey our sincere thanks to Bhaswar Chatterjee and Mujeeb Hasan
for their constant help during this work.

\end{document}